\documentclass[aps,prb,twocolumn,showpacs,superscriptaddress, groupedaddress]{revtex4-2}
\usepackage[english]{babel}
\usepackage{graphicx, color}
\usepackage{hyperref}
\definecolor{link}{rgb}{0.1,0.1,0.9}
\hypersetup{colorlinks=true,linkcolor=link,citecolor=link,urlcolor=link,linktocpage}
\usepackage{epstopdf}
\usepackage{color}
\usepackage{amsmath}
\usepackage{amssymb}
\usepackage{longtable}
\usepackage{mathtools}
\usepackage{float}
\usepackage{subfigure}

\setcitestyle{numbers,square}
\begin{document}

\title{Origin of magnetic anisotropy in La$_{1-x}$Sr$_{x}$MnO$_{3}$}
 
\affiliation
{School of Physical Sciences, Jawaharlal Nehru University, New Delhi 110067, India}
\author{Birendra Kumar}
\author{Harish Chandr Chauhan}
\author{Ajay Baro}
\author{Jyoti Saini}
\author{Ankita Tiwari}
\affiliation
{School of Physical Sciences, Jawaharlal Nehru University, New Delhi 110067, India}
\author{Mukesh Verma}
\affiliation
{Department of Physics, Central University of Rajasthan, Bandar Sindri 305817, India} 
\author{Yugandhar Bitla}
\affiliation
{Department of Physics, Central University of Rajasthan, Bandar Sindri 305817, India}
\author{Subhasis Ghosh}
\email{subhasis.ghosh.jnu@gmail.com}
\affiliation{School of Physical Sciences, Jawaharlal Nehru University, New Delhi 110067, India%
}

\begin{abstract}
Here, we report the origin of magnetic anisotropy in Sr doped infinite layer manganites La$_{1-x}$Sr$_{x}$MnO$_{3}$ ($ 0.125 \leq x \leq 0.400 $). The magnetic anisotropy is responsible for large difference in temperature dependence of field cooled and zero field cooled magnetization. Translational symmetry breaking in context of spins around the boundary between ferromagnetic (FM)-antiferromagnetic (AFM) region leads to FM-AFM interaction, and results in magnetic anisotropy (exchange anisotropy). Here, we propose that FM-AFM interaction around the boundary between FM clusters or domains in AFM background or between AFM clusters or domains in ferromagnetic background is reponsible for doping dependent nonmonotonic behavior and origin of magnetic anisotropy. 
\end{abstract}
\maketitle

\section{Introduction}
\hspace{6 mm} Manganites R$_{1-x}$A$_{x}$MnO$_{3}$ (R: rare$ - $earth ions, A: alkaline$ - $earth ions) have attracted great deal of interests as these compounds exhibit several fundamental phenomena, such as colossal magneto resistance (CMR)\cite{KAGAN20211, baldini2015origin, Tokura_2006,  PARASKEVOPOULOS2000118, PhysRevLett.92.157203, PhysRevB.51.14103,  PhysRevB.103.L161105, PhysRevB.105.035104, PhysRevB.62.1033, mahendiran1995}, magneto caloric effect (CME) \cite{SALAZARMUNOZ2022169787, Ram2018-we}, both second order and first order phase transitions (SOPT and FOPT), and metal-insulator transitions (MIT) \cite{Xia2020-tx}. Furthermore, manganites are actively being considered for future memory and magnetocaloric devices. Interest in manganites has been renewed after the observation of topological spin structures, such as stripes, bubbles, skyrmion in infinite layer, as well as in bilayer manganites \cite{GOBEL20211, PhysRevB.100.064429, Nagao2013-on, Kumar.Tiwari.2020, Kumar2021-sp, TIWARI2021168020, Tiwari2021-su, Wang2022-ku, Yu2014-if,  Gobel2019-mu}. All these novel magnetic properties are resulted from collective phenomena arising from the strong interplay among spin, charge, orbital, and lattice degrees of freedom in manganites. The infinite layer manganites La$ _{1-x} $Sr$ _{x} $MnO$ _{3} $  (La$^{3+}_{1-x}$Sr$^{2+}_{x}$Mn$^{3+}_{1-x}$Mn$^{4+}_{x}$O$^{2-}_{3}$) with $ 0.0 \leq x \leq 1.0 $ (IL$ - $LSMO$ -x $) are the member of Ruddelsden$ - $Popper series manganites \cite{Ruddlesden.1958}. IL$ - $LSMO$ -x $ manganites are stabilized in perovskite structures having centrosymmetric crystal structurecite \cite{PhysRevLett.71.2331}. In undoped manganite LaMnO$ _{3} $ (La$ ^{3+} $Mn$^{3+}$O$^{2-}_{3}$), the superexchange (SE) interaction \cite{MARKOVICH20141} between two Mn$ ^{3+} $ ions through O$ ^{2-} $ is responsible for ferromagnetism (FM) in MnO$ _{2} $ planes, and antiferromagnetic (AFM) ordering between MnO$ _{2} $ planes following Kanamori$ - $Goodenough formalism. However, the competing SE interaction in and perpendicular to MnO$ _{2} $ plane gives rise to AFM behavior. This is responsible for large magnetic anisotropy in LaMnO$ _{3} $ \cite{PhysRevB.54.15149}. Since, there is no hopping of charge carriers between singly occupied e$ _{g} $ state of Mn$ ^{3+} $ ions, LaMnO$ _{3} $ behaves as an antiferromagnetic insulator. When LaMnO$ _{3} $ is doped with Sr, there exists double exchange (DE) interaction \cite{MARKOVICH20141, PhysRev.118.141} between singly occupied e$ _{g} $ state of Mn$ ^{3+} $, and empty e$ _{g} $ state of Mn$ ^{4+} $. In DE mechanism, one electron hops from singly occupied e$ _{g} $ state of Mn$ ^{3+} $ to empty e$ _{g} $ state of Mn$ ^{4+} $ through O$ ^{2-} $ leading to enhanced FM and conducting nature in IL$ - $LSMO$ -x $. Hence, magnetic and conducting behaviours of IL$ - $LSMO$ -x $ depend on the doping concentration $ x $, contribution of SE, DE, and other effects. In addition to SE and DE interactions, doping plays important role in Jahn-Teller (JT) distortions, orbital$ - $ordering (OO) or charge$ - $ordering (CO)  \cite{PhysRevB.60.7006, Tokura_2006, PhysRevB.51.14103, PhysRev.118.141} in manganites. In the concentration range of $ 0.0 < x < 0.10 $, IL$ - $LSMO$ -x $ show canted antiferromagnetic (CAFM) insulating behaviour under JT distortion along c$ - $axis \cite{PhysRevB.51.14103, PhysRev.118.141}. This canting is based on the mechanism put forward by de Gennes \cite{PhysRev.118.141}. For $ 0.10 \leq x \leq 0.17 $, IL$ - $LSMO$ -x $, OO causes ferromagnetic and insulating behaviour due to phase separation \cite{PARASKEVOPOULOS2000118, PhysRevB.60.7006} which is believed to be responsible for CMR \cite{mahendiran1995, PARASKEVOPOULOS2000118, urushibara1995insulator, Tokura_2006} in manganites. And, for $ 0.17 < x < 0.40 $, IL$ - $LSMO$ -x $ show ferromagnetic and conducting behaviours in absence of both the JT$ - $distortion and the OO. As discussed above, there exits both SE and DE interactions in doped IL$ - $LSMO$ -x $ ($ 0.0 < x < 1.0 $). Hence, the magnetic properties and the respective contribution in deciding the overall magnetic properties depended on doping concentration $ x $. The ratio of Mn$ ^{4+} $ to Mn$ ^{3+} $ increases with $ x $, $ i.e. $, there is enhancement in DE contribution while decrease in SE contribution, which leads to increase in FM ordering and conducting nature up to a critical doping, beyond which IL$ - $LSMO$ -x $ starts behaving as AFM and insulating due to over$ - $doping that leads to emergence of A$ - $type AFM in which the spins lie in the MnO$ _{2} $ plane resulting stripes and other topological spin configurations \cite{PhysRevLett.78.4253}. The increase in both the transition$ - $temperature (T$ _{C} $) and conducting$ - $behaviour with Sr doping, as presented in the phase diagram (supplementary materials), justifies the tuning of DE and SE. Generally, temperature$ - $dependent magnetization (M$ - $T) of FM and AFM materials in field$ - $cooled (FC) and zero$ - $field$ - $cooled (ZFC) modes show bifurcation below the transition point. It is generally believed that the bifurcation is one of the characteristic features for the existence of magnetic anisotropy \cite{kumar1998origin, joy1998}. Higher the magnitude of bifurcation the greater will be the magnetic anisotropy. 

The bifurcation (Fig. 1) in the M$ - $T for IL$ - $LSMO$ -0.125 $ is a typical signature for the magnetic anisotropy. Similar bifurcation and cusp, in FC and ZFC modes, have been observed in spin glass, superparamagnetic, ferromagnetic and ferrimagnetic materials and always speculated that some kind of anisotropy is responsible for this bifurcation. Anisotropy results from (i) spin-orbit coupling (SOC): in non-centrosymmetric systems, such as FeGe, MnSi and Cu$ _{2} $OSeO$ _{3} $, induced SOC or antisymmetric exchange Dzyaloshinskii-Moriya interaction (DMI) gives rise to anisotropy \cite{muhlbauer2009, tonomura2012, PhysRevLett.107.127203, Yu2011, PhysRevLett.110.077207, seki2012, PhysRevB.85.220406, PhysRevB.91.224408, PhysRevB.100.165143, PhysRevLett.128.015703}, however manganites are centrosymmetric, so SOC driven anisotropy can be ruled out. (ii) Stress anisotropy(magnetostriction):  stress driven anisotropy is known as magnetoelastic anisotropy and generally observed in crystal with sublattices or when the crystal is under external stimulai. So, magnetoelastic anisotropy is absent in the manganites. (iii) Shape anisotropy: due to dipole-dipole interaction the demagnetizing field will not be same in all directions resulting multiple easy axis. However, shape anisotropy is absent in polycrytalline sample as grains or domains are distributed evenly in all directions.  (iv) Surface and interface induced anisotropy: the broken symmetry at surfaces and interfaces of magnetic thin films and multilayers often induces some anisotropy in the system. In the bulk manganites, the surface and interface anisotropy are absent. (v) Unidirectional anisotropy: this anisotropy is due to exchange bias in the presence of both FM and AFM, but this anisotropy can be ruled out in polycrystalline samples. Finally, (vi) anisotropy due to FM-AFM frustration: the strong anisotropy and large bifurcation between FC and ZFC magnetization observed in spin-glass systems is due to frustration driven by FM- AFM completion and degeneracy in ground state.  The frustration leads to random orientation of spins in the dilute magnetic system. The systematic pattern of CMR observed around T$ _{c} $ and absence of exchange bias in manganites \cite{mahendiran1995, PARASKEVOPOULOS2000118, urushibara1995insulator, Tokura_2006} rules out the spin glass like ground state in manganites in spite of presence of both FM and AFM in doped manganites as discussed above.  

In absence of all known anisotropy in manganites, the natural question comes-what is microscopic origin of anisotropy in manganites?  As discussed before, the Sr doping in LaMnO$ _{3} $ critically decides the magnetic and conducting properties of manganites by controlling the doping dependent SE and DE interactions leading to coexistence of FM and AFM, and their dominancy on each other.  It is surprising that this basic issue has been ignored so far in the context of anisotropy responsible for bifurcation in ZFC and FC magnetization. 
 
\begin{figure}
	\centering{}
	\includegraphics[width=\linewidth]{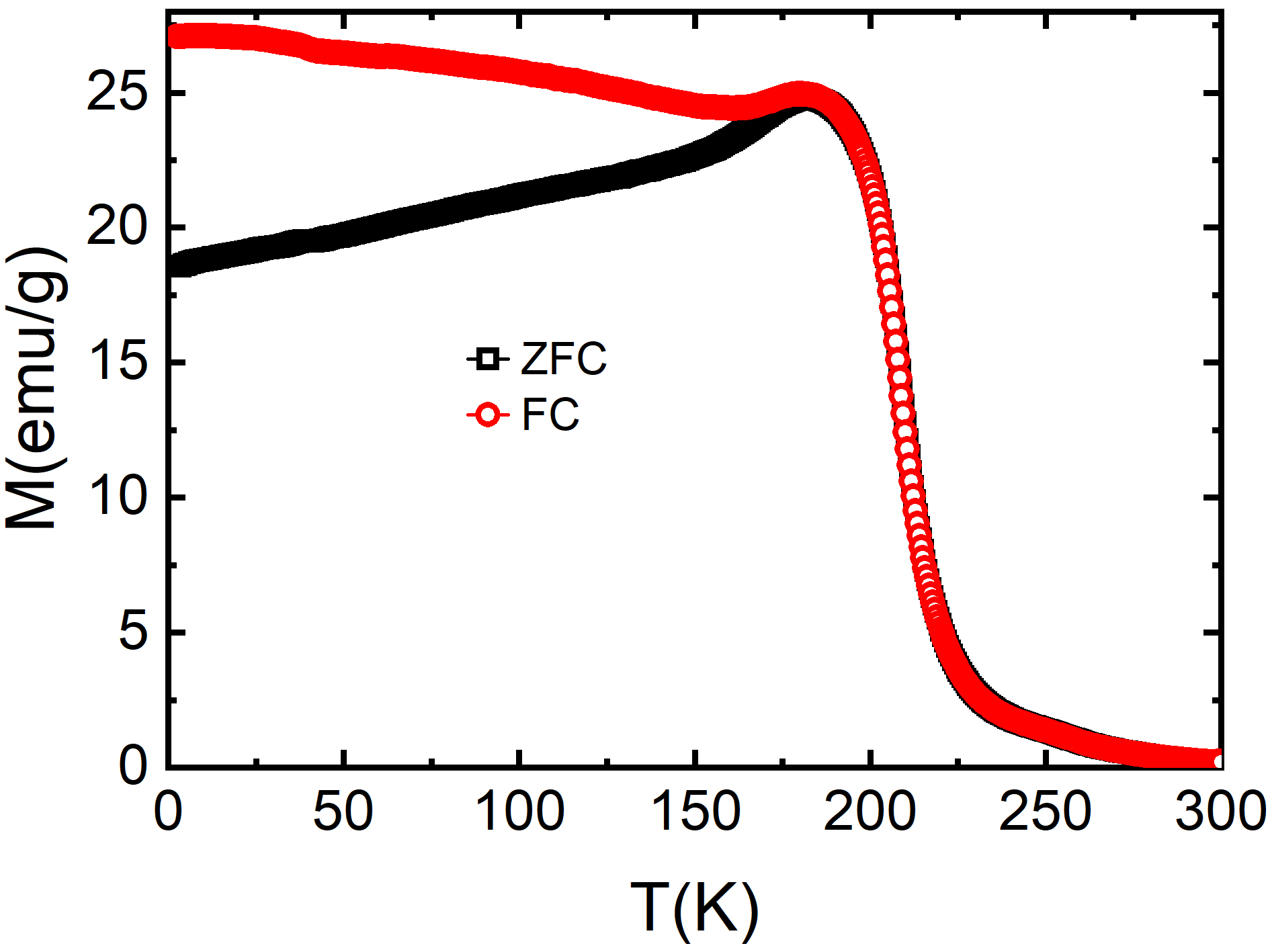}
	\caption{Temperature$ - $dependent magnetization for IL-LSMO-0.125, obtained in FC and ZFC warming mode in presence of applied field 10 mT. Bifurcation, \textit{i.e.}, irreversibility, in FC and ZFC, is observed below T$_C$.}
	\label{MT}
\end{figure}

From Fig. \ref{MT} two observations can be made: (1) relatively large anisotropy (difference between FC and ZFC) in dilute manganites and (2) apparent antiferromagnetic nature of MT in ZFC mode. Here, we show that the possible reason for magnetic anisotropy in manganites may be the interaction between the spins of FM-phase and AFM-phase around the boundary of FM-AFM region. The observed CMR in the manganites \cite{mahendiran1995, PARASKEVOPOULOS2000118, urushibara1995insulator, Tokura_2006} confirms FM- and AFM- phase separation, and therefore there exist interfaces or the boundaries between FM and AFM domains or clusters. Hence, whether ferromagnetic properties of manganites   is predominantly determined by  FM cluster in AFM domain or AFM cluster in FM domain, is decided  by doping, and must be separated by a critical doping concentration, $ x_{c} $.  In both cases, the non-symmetric orientation of spins around the FM-AFM interface may results in exchange anisotropy in the manganites. To establish the importance of this exchange anisotropy on the bifurcation observed in magnetization in doped manganites, we present experimental investigation on the dependence of the degree of bifurcation on doping range of $ 0.125 \leq x \leq 0.40 $.  The convergence of anisotropy determined using different methods and nonmonotonic dependence on degree of anisotropy on doping concentration can be attributed to the coexistence of FM-AFM as the microscopic origin of anisotropy in manganites.  

\section{Experimental Details}
\hspace{6 mm} IL-LSMO-x samples with varying values of x (0.125, 0.175, 0.200, 0.225, 0.300 and 0.400) have been prepared using the standard solid-state reaction route \cite{KumarTiwari2020}. The detailed procedure for sample preparation is given in the supplementary material (SM)\cite{SM}. The crystal structure of the samples was determined by performing X-ray diffraction (XRD) measurements at room temperature using a Rigaku X-ray diffractometer with Cu-K$ _{\alpha} $ line. M$ - $T measurements were carried out on all the samples under an applied magnetic field of 10 mT. These measurements provide valuable insights into the magnetic properties of the samples.
The M-T measurements were performed in two modes: zero-field cooled (ZFC) and field cooled (FC) warming modes. In the ZFC mode, the sample is cooled without any magnetic field, and then the magnetization is measured as the temperature is increased. In the FC mode, the sample is cooled while a magnetic field is applied, and then the magnetization is measured as the temperature is increased. The four-quadrant field$ - $dependent magnetization (M$ - $H) measurements were conducted at a temperature of 10 K for all the samples. These magnetic measurements were carried out using the VSM (Vibrating Sample Magnetometer) option of a physical properties measurement system, specifically the Quantum Design system.  

\section{Results and discussion}
\subsection{{Structural analysis}}
\begin{figure}
	\centering{}
	\includegraphics[width=\linewidth]{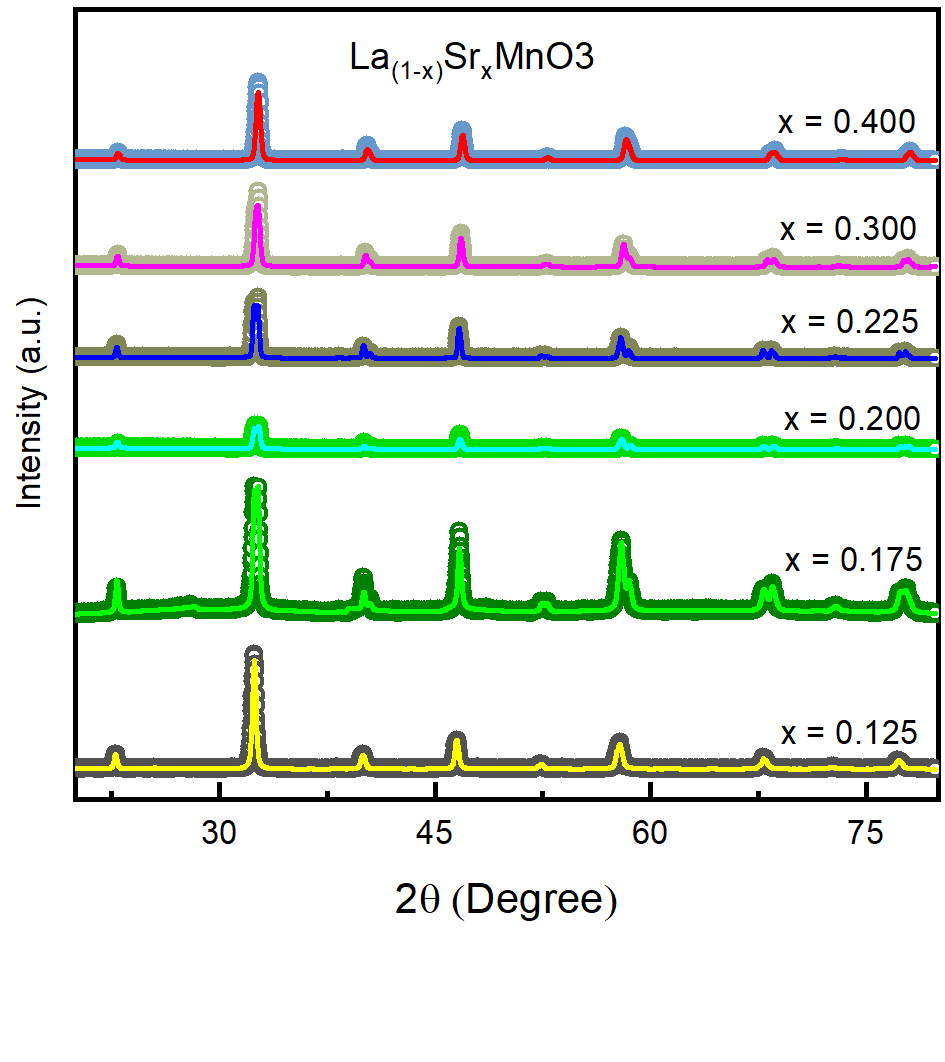}
	\caption{Room temperature XRD of IL-LSMO-$x$ ($x=0.125, 0.175, 0.200, 0.225, 0.300,$ and $0.400$). The solid circles and lines represent the raw and refined data, respectively. The XRD pattern (raw and refined) of all the samples has been shifted along the intensity axis.}
	\label{XRD}
\end{figure}

\hspace{6 mm} The phase purity of the IL$ - $LSMO$ - $x samples has been confirmed through Rietveld refinement of the XRD patterns, as shown in Fig. \ref{XRD}. The absence of impurity peaks indicates the formation of single-phase IL$ - $LSMO$ - $x samples. The crystal structure of IL-LSMO-x varies with different values of x. For x = 0.125, the crystal structure is orthorhombic. However, it transforms into a rhombohedral structure for x = 0.175. In the range of $0.200\le x \le 0.400$, the crystal structure of IL$ - $LSMO$-x$ is trigonal. The crystal parameters, including lattice constants ($a, b,c$) and volume $V$, for all the samples can be found in the reference \cite{SM}.

\subsection{Magnetic properties}
\subsubsection{Temperature-dependent magnetization}
\begin{figure*}
	\centering{}
	\includegraphics[width= \linewidth]{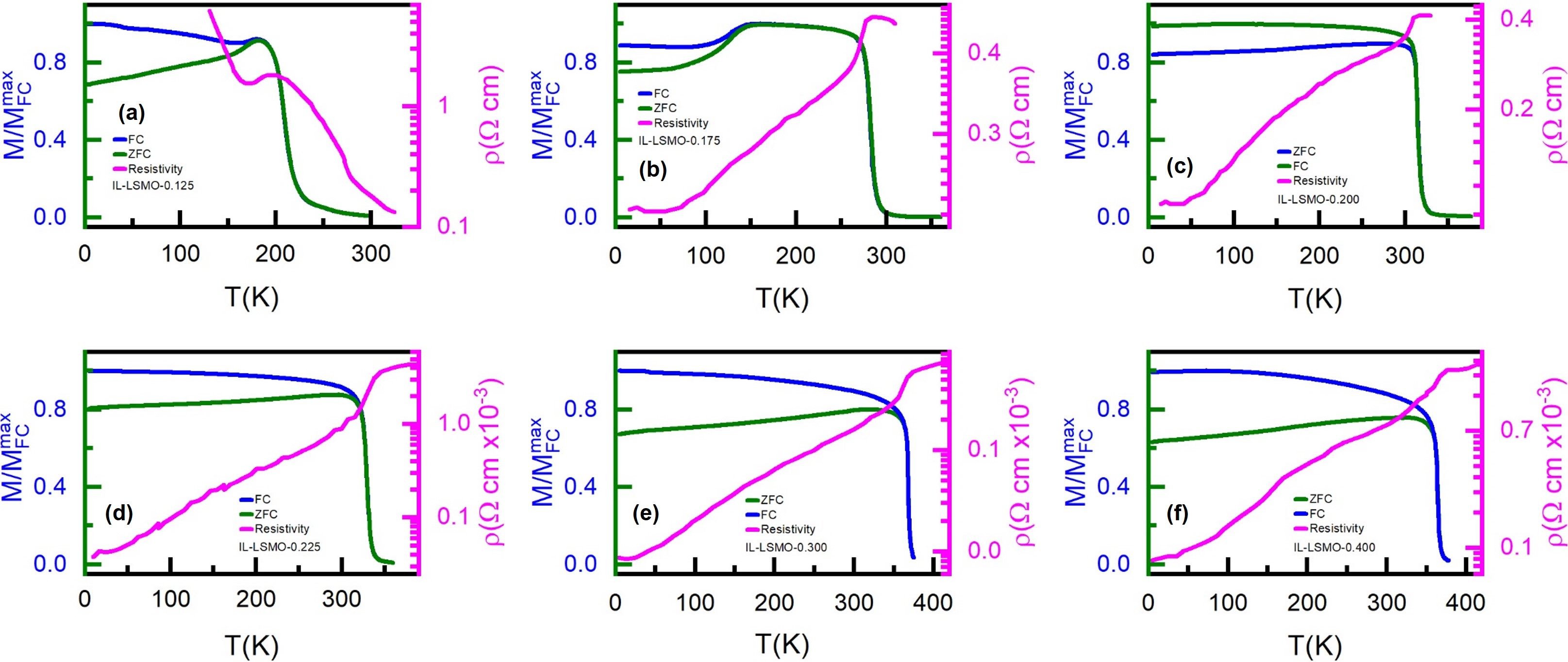}
	\caption{Temperature$ - $dependent normalized magnetization (${M/M^{max}_{FC}}$$ - T$) in both FC (blue solid lines) and ZFC (green solid lines) modes of measurement and resistivity (mangeta solid lies) for (a) IL-LSMO-0.125, (b) IL-LSMO-0.175 (c) IL-LSMO-0.200, (d) IL-LSMO-0.2250, (e) IL-LSMO-0.300, and (f) IL-LSMO-0.400. The resistivity data have taken from the reports \cite{PhysRevB.51.14103, Dabrowski1999StructurepropertiesPD, PhysRevB.68.134428}}
	\label{FC-ZFC}
\end{figure*}
\hspace{6 mm} Figures \ref{FC-ZFC}(a) to \ref{FC-ZFC}(f) show M$ - $T isofields and resistivity of IL$ - $LSMO$-x$ measured in FC and ZFC warming-mode. All the M$ - $T isofields have been obtained at 10 mT and the resistivities have been obtained from references \cite{PhysRevB.51.14103, PhysRevB.60.7006, PhysRevB.68.134428}. It is found that $T_C$ increases with Sr doping up to $x = 0.300$, while a decrease in $T_C$ is observed for $x = 0.400$. The trend of increasing $T_C$ with Sr doping until $x=0.300$, followed by a decrease at $x=0.400$, is consistent with previous reports \cite{PhysRevB.60.7006, PhysRevB.51.14103, PARASKEVOPOULOS2000118}. Based on the DE and SE mechanisms discussed earlier, the explanation for this trend can be understood as follows:
In LaMnO$_3$, the SEI between Mn$^{3+}$ ions through O$^{2-}$ anions promotes AFM behavior. However, with Sr doping in IL-LSMO-$x$, the induced DEI emerges due to the presence of $Mn^{3+}-O^{2-}-Mn^{4+}$ configurations. This DEI favors FM behavior.
The increase in $T_C$ for $x\leq0.300$ indicates that the contribution of the DEI is more significant than that of the SEI. As Sr doping increases, the FM clusters mediated by Mn$^{3+}-O^{2-}-Mn^{4+}$ configurations become more pronounced within the AFM matrix. This enhancement of the FM clusters leads to an increase in $T_C$. However, at $x=0.400$, a decrease in $T_C$ is observed. This change suggests that the balance between the DEI and SEI mechanisms is shifting. It is possible that the increasing Sr doping beyond $x=0.300$ starts to disrupt or suppress the FM clusters mediated by Mn$^{3+}-O^{2-}-Mn^{4+}$ configurations, leading to a decrease in $T_C$. Overall, the observed trend in the change of $T_C$ in IL-LSMO-$x$ materials with Sr doping is consistent with the interplay between the DEI and SEI mechanisms, where the contribution of the DEI becomes more dominant up to a certain Sr doping level, resulting in an increase in $T_C$.
For $x<0.200$, AFM behavior is observed at low temperatures due to the effective SEI, which leads to the formation of FM clusters within the AFM background. This results in the presence of AFM-FM interfaces at low temperatures in IL-LSMO-$x$. As the temperature increases below the transition point, the strength of the SEI weakens, leading to an increase in the size of the FM clusters due to a percolation phenomenon. This behavior is consistent with percolation theory \cite{PhysRevLett.87.277202}. Above a critical temperature $T^*$, the strength of the SEI becomes very low, favoring FM behavior due to the effective DEI. This FM behavior is observed in the high-temperature regime.

The transport behavior of IL$ - $LSMO-$x$ materials also supports the increase in $T_C$ in a similar fashion. 
The resistivity versus temperature curves show insulating behavior below $T^*$, followed by a conducting regime between $T^*$ and $T_C$, and again insulating behavior above $T_C$ \cite{urushibara1995insulator, PhysRevB.60.7006}. In the range $0.125\leq x\leq 0.200$, the percolation phenomenon leads to the destruction of FM-AFM interfaces and an increase in the metal-insulator transition temperature ($T_{MI}\equiv T_C$) with doping. For $0.200\leq x\leq 0.400$, the entire temperature regime below $T_C$ exhibits FM and conducting behavior. This can be explained as follows \cite{PhysRevLett.87.277202}: for $0\leq x<0.200$, there are separated FM clusters within the AFM background. At a critical Sr doping level ($x\simeq 0.200$), some of the FM clusters become connected, leading to conduction. For $0.200<x\leq 0.400$, the background itself becomes FM, while there are still AFM clusters present. 
The observations suggest that above $T^*$, IL-LSMO-$x$ materials become FM due to the percolation of FM clusters, which destroys the FM-AFM interfaces. The metal-insulator transition temperature (T$ _{MI} \equiv T_C $) increases with doping in the range $0.125\leq x\leq 0.300$. At and above $x=0.200$, FM behavior and conduction are observed throughout the temperature regime below $T_C$.

The FC and ZFC M$ - $T isofields show significant bifurcation below $T_C$ as shown in Fig. \ref{FC-ZFC}. The magnitude of bifurcation first decreases with $ x $ for $0.125\le x \le 0.200$ [Figs \ref{FC-ZFC}(a) to \ref{FC-ZFC}(c)], and then increases for $x>0.200$ [Figs \ref{FC-ZFC}(c) to \ref{FC-ZFC}(f)]. Generally, the bifurcation and cusp are observed in spin$ - $glass, ferromagnetic and ferrimagnetic materials. In spin-glass materials, bifurcation is observed due to frustration caused by competing AFM and FM exchange$ - $interactions, deformed lattices, and random orientation of ions\cite{RevModPhys.58.801, mydosh1993spin} while magnetocrystalline anisotropy is responsible for the formation of cusp and bifurcation under weak magnetic field\cite{kumar1998origin, joy1998} in FM and ferrimagnetic oxides.
\begin{figure}[h]
	\centering{}
	\includegraphics[width=\linewidth]{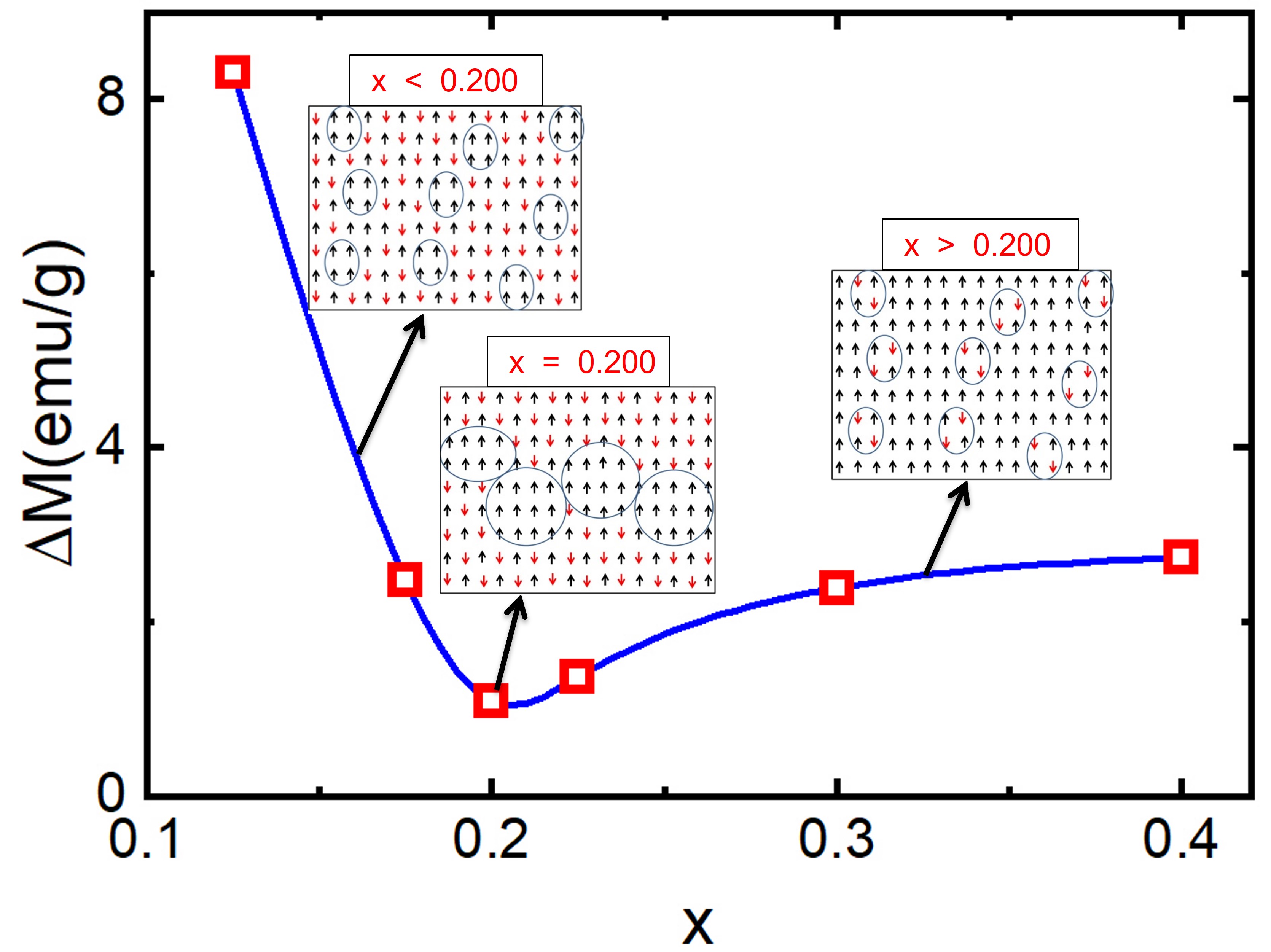}
	\caption{The difference in magnetizations for FC and ZFC taken at 10 K i.e. $ \Delta M $ vs. $x$ plots for IL-LSMO-$x$ for $x=0.125, 0.175, 0.200, 0.225, 0.300$, and $0.400$. The data for all the compositions have been recorded at 10 mT. The schematic representation of cluster model: FM clusters in AFM matrix for $0.125\le x < 0.200$ and AFM clusters in FM matrix for $0.200\le x \le 0.400$.}
	\label{fig2}
\end{figure}
The systematic observed CMR behavior around T$ _{c} $, particularly at high magnetic fields \cite{mahendiran1995, PARASKEVOPOULOS2000118, urushibara1995insulator, Tokura_2006}, and the negligible exchange bias (EB) in manganites suggest that they are not spin-glass systems. Therefore, the bifurcation observed in the magnetization curves of the IL-LSMO-$x$ ($0.125 \leq x \leq 0.400$) manganites is attributed to factors other than spin-glass characteristics. The bifurcation in magnetization for FC and ZFC modes may be explained as follows: In the absence of a magnetic field, as the temperature of a magnetic material decreases, the thermal energy decreases, resulting in a reduction of thermal excitation. Consequently, the spins in the material freeze and align in a direction determined by the lattice and hybridized orbital orientation, influenced by the degeneracy at that temperature. However, in the presence of a magnetic field, as the temperature decreases, the thermal excitation decreases, causing the spins to freeze and align in the direction of the magnetic field. The applied magnetic field also reduces the degeneracy that supports ferromagnetism. As a result, M$ - $T curves differ between the FC and ZFC modes, and the difference between the FC and ZFC magnetization curves decreases as the applied field increases up to a critical field. Beyond this critical field, the material becomes field-polarized ferromagnetic. 
As the magnitude of bifurcation between M$ _{FC} $ and M$ _{ZFC} $ is dependent on both the magnetic anisotropy and temperature\cite{kumar1998origin, joy1998}, so, higher the anisotropy the greater will be bifurcation between FC and ZFC modes of magnetization\cite{kumar1998origin, joy1998}. Thus, magnetic irreversibility $\Delta$M, i.e., the difference M$_{FC}$$ - $M$_{ZFC}$ gives an indirect signature for the existence of magnetic anisotropy. From the plot of $\Delta M$ vs. $x$ (Fig. \ref{fig2}) for IL-LSMO-$ x $ ($0.125 \leq  x \leq 0.400$)), it can be argued that the anisotropy decreases with doping for $0.125\le x \le 0.200$ and increases for $0.200\le x \le 0.400$. This unusual variation of bifurcation may be explained by cluster model as follows: since doped manganites exhibit both SE and DE interactions that leads to coexistence of FM and AFM mixed phase in the form FM (or AFM) clusters in the matrix of AFM (or FM) phase doping upon the doping concentration. In the lower doping concentration range ($0.125\le x < 0.200$), FM (separated) clusters exist in AFM matrix with connected clusters (may be comparable size of FM and AFM clusters or domain) at $ x = 0.200 $ while in higher doping concentration range ($0.200\le x \le 0.400$) AFM clusters are in the FM matrix. This cluster model will also explain the conducting behavior when FM clusters are connected at  $ x \geq 0.200 $. The cluster model is schematically represented in Fig. \ref{fig2} 

\subsubsection{Field-dependent magnetization}
\begin{figure}[h]
	\centering{}
	\includegraphics[width= \linewidth]{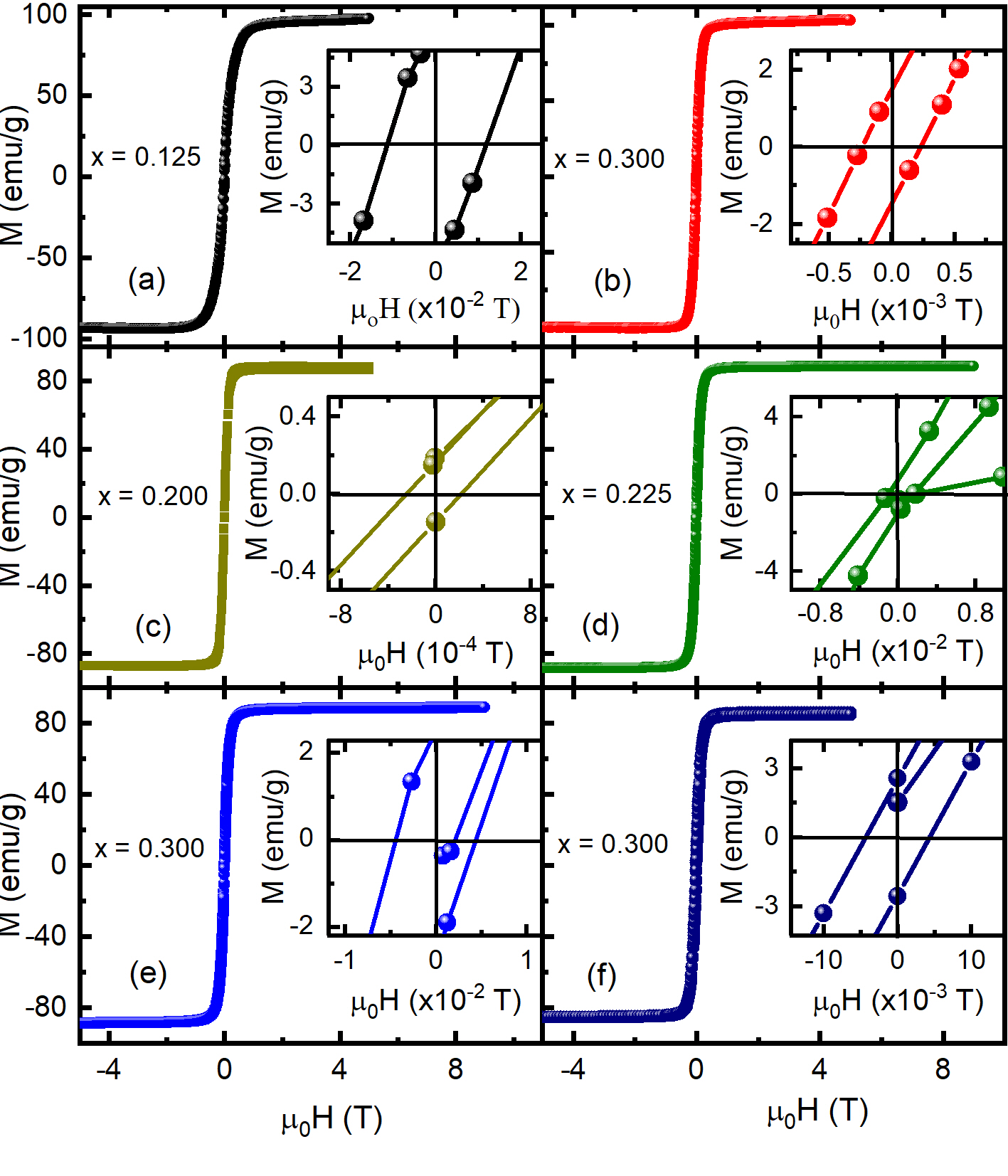}
	\caption{Field$ - $dependent magnetization (M$ - $H) loops against x taken at 10 K for: (a) IL-LSMO-0.125, (b) IL-LSMO-0.175, (c) IL-LSMO-0.200, (d) IL-LSMO-0.225, (e) IL-LSMO-0.300, and (f) IL-LSMO-0.400. The insets are expanded view of the magnetic hysteresis with coercive fields for all the composition.}
	\label{fig5}
\end{figure}

\hspace{6 mm} Figure \ref{fig5} showing four quadrants M$ - $H for IL-LSMO-$x$ at 10 K, exhibits significantly small hysteresis (soft FM nature). The observation of magnetic hysteresis in M$ - $H curves, is the signature for the existence of magnetic anisotropy in the material. In general, the formation of domain wall causes magnetic hysteresis to exist in a material \cite{Skomski2008Magnetism}. The coercivity is proportional to the width of the domain wall. The anisotropy causes the formation of domain walls in which spins are oriented in different directions from the magnetic easy axis. Quantitatively, the magnetic anisotropy is determined by the relation \cite{PhysRev.105.904, PhysRev.96.1335, PhysRevB.53.14415}
\begin{equation}
K_{an} = \frac{M_{s}H_{c}}{2},
\label{Kan}
\end{equation}
where $M_{s}$ and $H_{c}$ are the saturation magnetization, and coercive field, respectively. The variation of $H_c$ (Fig. \ref{fig6}) is similar to the variation of $ \Delta M $ vs. $x$ (Fig. \ref{fig2}) for IL-LSMO-$x$, \textit{i.e.}, $H_c$ first decreases for $x \le 0.200$, and increases for $x>0.200$. The similar dependence of $H_c$ and $K_{an}$ on $ x $ can be speculated that they have common origin.

\subsubsection{Correlation between magnetic anisotropy from M-H and M-T}
\begin{figure}[h]
	\centering{}
	\includegraphics[width=\linewidth]{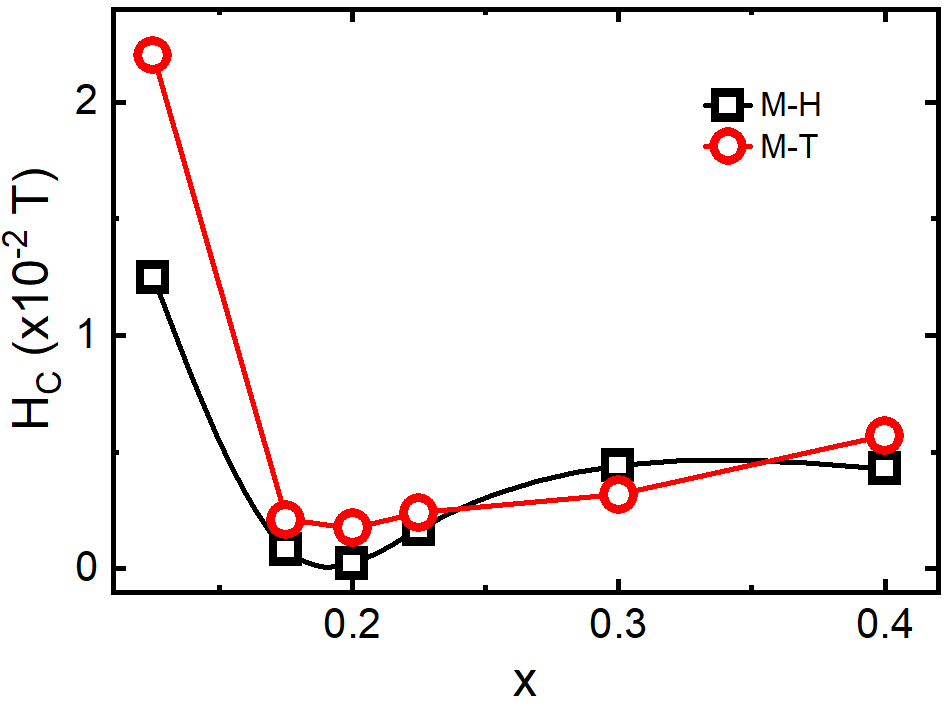}
	\caption{Variation of $H_{c}$ with doping $ x $ at 10 K for L-LSMO-$ x $ ($0.125 \leq x \leq 0.400$). H$_{c}$$ - $$ x $ plots constructed from M$ - $H (curve with square symbol) and from M-T (curve with circle symbol). The square and circle symbols represent data points, and the solid curves are the splines for eye guide.}
	\label{fig6}
\end{figure}

\hspace{6 mm} The M$ - $H at 10 K for IL-LSMO-$ x $ ($ x $ = 0.125, 0.175, 0.200, 0.225, 0.300 and 0.400) shows the magnetic saturation at high magnetic field (Fig. \ref{fig5}). This means that the dominating behavior of all samples below their corresponding transition point is FM. The M$_{s}$ has been calculated from Fig. \ref{fig5}, given in table \ref{table3}, which shows that the M$_{s}$ first decreases with $ x $ for $ x \leq $ 0.175 then increases with further increase in $ x $ for $ x > 0.175 $, and finally decreases at x $ = $ 0.400. This $M_{s}$ variation is similar to the composition dependent electrical transport feature for IL-LSMO-$ x $ ($ 0.125 \leq x \leq 0.400 $) as discussed above. 
This kind of transport behavior is well explained by cluster model, $ i.e, $ the existence of FM clusters in the background of AFM phase or AFM clusters in FM background \cite{PhysRevLett.87.277202}. The magnetic coercive field, H$_{c}$, for IL-LSMO-$ x $ has been calculated from M$ - $H (Fig. \ref{FC-ZFC}) and M$ - $T curves. The values of H$_{c}$ have been determined directly from M$ - $H (Fig. \ref{fig5}) and indirectly from M$ - $T using the relation    
\begin{equation}
\frac{M_{FC}}{H_{app} + H_{c}} \approx \frac{M_{ZFC}}{H_{app}},
\end{equation}
where $H_{app}$ is the applied magnetic field (10 mT for all the samples). The H$_{c}$ calculated from both M$ - $H and M$ - $T at 10 K for all IL-LSMO-$ x $ ($ x $ = 0.125, 0.175, 0.200, 0.225, 0.300 and 0.400) are given in table \ref{table3}, and shown in Fig. \ref{fig6}. The H$_{c}$ values determined from M$ - $H and M$ - $T measurements are in close agreement and exhibit dependence on the doping concentration $ x $.     
From Fig. \ref{fig6}, it is clear that $H_{c}$ first decreases with Sr$ - $doping for $0.125 \leq x \leq 0.200 $ then increases with further increment in $ x $ for $0.200 \leq x \leq 0.300 $, and finally decreases at $x = $ 0.400. This feature of $H_{c}$ $ vs. x $  for $ x \leq 0.200 $ can not be explained by only the competing contribution of SE and DE interactions because with increases in $ x $ the contribution of DE increases for the range $ 0.125 \leq x \leq 0.300 $, and decreses for $x  =  0.400$. Thus, in the composition range $ x \leq 0.200 $, $H_{c}$ is controlled by not only SE and DE interactions but also by other parameters such as JT distortion, OO, phase separation or segregation, and magnetic anisotropy. Based on previous reports, phase diagram have been constructed and presented in supporting material\cite{SM} with is in consistent with our results for all the compositions. 
\begin{table}
\centering 
\caption {M$ _{s} $, H$_{c}$ and K$ _{an} $ at temperature 10 K for different compositions $ x $.}	 
	\begin{tabular}{| c | c | c |c |c | } \hline 
		$ x $ & Ms & Hc from M-H  & Hc from M-T & K$ _{an} $ \\ \hline
		Unit & emu/g & (T) & (T)  & (emu T/g) \\ \hline
		0.125 & 97.4979 & 0.01246  & 0.0220   & 0.6079 \\ \hline
		0.175 & 82.9459 & 0.00235  & 0.0021   & 0.0975 \\ \hline
		0.200 & 88.2676 & 0.00023  & 0.0017   & 0.0103 \\ \hline
		0.225 & 88.5706 & 0.00170  & 0.0024   & 0.0753 \\ \hline
		0.300 & 88.7667 & 0.00443  & 0.0048   & 0.1966 \\ \hline
		0.400 & 86.0778 & 0.00429  & 0.0057   & 0.1846 \\ \hline
	\end{tabular}
	\label{table3}
\end{table}
\subsection{Discussion}
\begin{figure}[h]
	\centering{}
	\includegraphics[width= 6 cm]{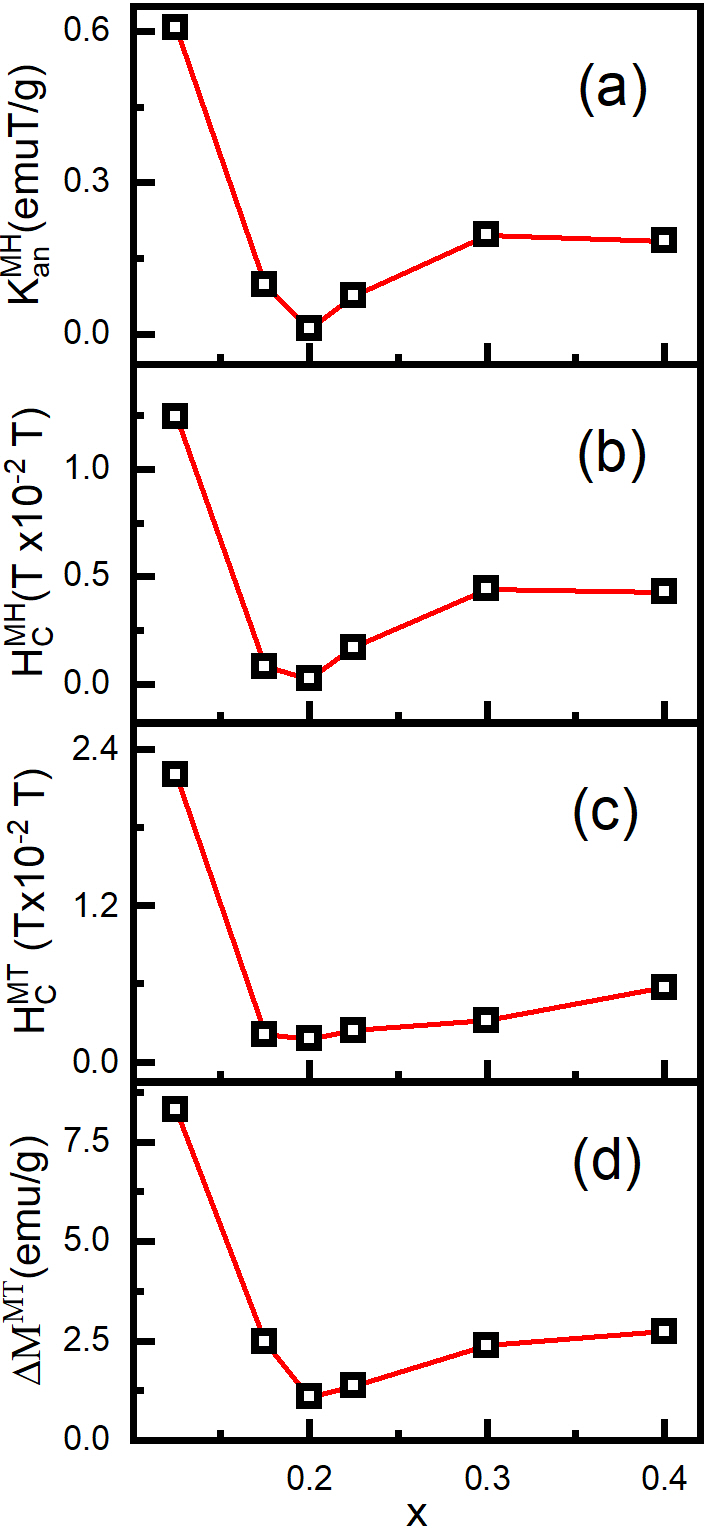}
	\caption{Magnetic anisotropy study of IL-LSMO-$ x $ ($ x $ = 0.125, 0.175, 0.200, 0.225, 0.300, and 0.400) at 10 K: (a) $K_{an}$ determined from M-H against doping concentration $ x $, (b) H$ _{c} $ versus $ x $ plot using M-H, (c)  H$ _{c} $ versus $ x $ plot using FC-ZFC, and (d) difference in magnetization of FC and ZFC modes. The square symbol represents data points and the solid curve is the spline for eye guide.}
	\label{fig7}
\end{figure}
 
\hspace{6 mm} The magnetic anisotropy is determined by using the anisotropic energy $E _{an} $ \cite{PhysRev.105.904, PhysRev.96.1335, PhysRevB.53.14415}:
\begin{equation}
E_{a} \approx K_{an}Sin^{2}\theta,
\label{Kan-1}
\end{equation}
where $\theta$ is the angular position of magnetization, $M$, from the easy axis, and $K _{an} $ is the anisotropy constant which quantitatively determines the extent of magnetic anisotropy. The $K _{an} $ depends on both $M_{s}$ and $H_{c}$ (Eq. \ref{Kan}). Hence, $K _{an} $ can be determined if both $M_{s}$ and $H_{c}$ are known. For IL-LSMO-x ($ 0.125 \leq x \leq 0.400 $), $M_{s}$ and $H_{c}$ (Fig. \ref{fig5}) have been calculated from their respective M-H taken at 10 K, and then $K_{an}$ has been calculated using Eq. \ref{Kan}.
Figure \ref{fig7}(b) and table \ref{table3} show the doping dependent variation of $K_{an}$ which is similar to that of the variation in H$_{c}$ (Fig. \ref{fig6}) and $ \Delta M $ (Fig. \ref{fig2}) with $ x $ . From the comparative study (Fig. \ref{fig7}) we have observed that (i) K$ _{an} $ determined from M$ - $H, (ii) H$ _{c}$ determined from M$ - $H and M$ - $T (FC-ZFC), and (iii)  $ \Delta M  $ determined from FC$ - $ZFC, shows similar nonmonotonic behavior that manifests the existence of competing FM$ - $ AFM interactions in the manganites. The behavior is similar for K$ _{an} $, H$ _{c} $ and $ \Delta M  $ for range 0.125 $\leq$ $ x $ $\leq$ 0.300. As discussed above both  $ \Delta M  $ and  H$ _{c}$ are the indicator for the existence of magnetic anisotropy, so the Fig. \ref{fig7} is the comparative study of magnetic anisotropy in different fashion.

As discussed, $K _{an} $ decreases with doping $ x $ for $ x \leq 0.200 $, and then increases with further increment in $ x $ for $x > 0.200$. This means the magnetization of IL-LSMO-$ x $ for $ x \leq 0.200 $ is nonuniform due to the existence of phase separation (FM clusters or domains in AFM background) while for $ x \geq 0.200 $ the nonunifomity increases due to presence of AFM clusters in the background of FM. The nonuniformity in magnetization decreases with doping for $ x \leq 0.200 $ while for the range $ 0.200 < x \leq 0.300$ the nonuniformity increases as doping changes with a minimum nonuniformity at $ x = 0.200 $. We have discussed in introduction, an anisotropy or an interaction in a system, is a result of a symmetry breaking, therefore, we must investigate which kind of symmetry is broken that induces anisotropy in IL-LSMO-$ x $. What is the interaction due to the symmetry breaking? The IL-LSMO-$ x $ manganites are in the form of layering with 3D system (no shape anisotopy as discussed in the introduction part). 
The possible symmetry breaking in the manganites is translational with respect to spin orientation about the boundary of FM clusters (or FM domains) in the AFM background, or AFM clusters in the background of FM. This induces FM-AFM interaction across the boundary which leads to exchange anisotropy. We explain the observed variation of magnetic anisotropy as a function of doping ($ x $) for IL-LSMO-$ x $ using combination of cluster model and exchange anisotropy as follows: for the doping range $ x \leq 0.200 $ as $ x $ increases the size of FM clusters increases (i.e., number of clusters decreases) and so, the size of AFM region decreases, as a result, the boundary contribution between AFM and FM regions decreases which leads to decrease in the exchange anisotropy. At $ x = 0.200 $, the size of FM domains (or clusters) become comparable (nearly equal) to the size of AFM domains (or clusters) as a result, the boundary between FM region and AFM region is minimum, and hence the magnetic anisotropy is minimum. The conducting behavior of IL-LSMO-$ 0.200 $ (Fig. \ref{FC-ZFC} and Fig. 2 of \cite{SM}) justifies that the FM clusters must be connected. Based on the cluster model, it can be conceived that for the doping range $ 0.200 \leq x $ $\leq$ 0.300, there is appearance of AFM clusters in FM background which is opposite to the case for doping range $ x < 0.200 $. Thus, for $ x $ $ < 0.200 $ the FM clusters are not connected leading to insulating behavior, and for $ x $ $ \geq 0.200 $ the FM domains are connected leading to conducting behavior. For the doping range $ 0.200 \leq x $ $\leq$ 0.300, the size of FM clusters or domain increases while the size of AFM clusters decreases, and hence the number of AFM clusters increases (opposite behavior to doping range $ x \leq 0.200 $). As a result, the boundary contributions between FM and AFM regions increases which leads to increase in the exchange anisotropy. At $ x = 0.400 $, the size of FM domains again decreases while the size of AFM clusters increases because the spins changes direction i.e., the AFM contribution changes from CE-type AFM to A-type AFM which leads to form some short of magnetic stripes with spins at some angle ($ 0 ^{0} < \theta < 90 ^{0} $) from the MnO$_{2}$ \cite{PhysRevLett.78.4253}. The size and connectivity of FM and AFM clusters are schematically represented in Fig. \ref{fig2} which visualize the transport and magnetic behaviors of IL-LSMO-$ x $. The schematic figure also shows that the boundary between FM and AFM region is lowest for $ x $ $ = 0.200 $ because the FM and AFM domains are of comparable size while for $ x < 0.200 $, and $ x \geq 0.200 $, AFM domains size is higher and lower than the size to that of FM domains, respectively. This leads to the minimum magnetic anisotropy at $ x $ $ = 0.200 $. Thus, in the doping range $ x $ $ \geq 0.200 $ and $ x < 0.200 $, magnetic anisotropy is due to FM-AFM exchange interaction across the boundary of FM and AFM regions. 
 
\section{Conclusions} 
\hspace{6 mm} Single crystal phase characterized by XRD of all the IL-LSMO-$ x $ ($ x $ = 0.125, 0.175, 0.200, 0.225, 0.300 and 0.400) have been prepared by standard solid state reaction. The crystal structure changes from orthorhombic to rhombohedral to trigonal with doping from $ x = 0.125$ to $ x = 0.400 $. In all the samples magnetic phase transition have been observed with increasing T$ _{c} $ from 208 K to 368 K for doping range $ 0.125 \leq x \leq0.300 $, however, a decrease in T$ _{c} $ at $ x = 0.400 $. Bifurcation between M$ - $Ts for FC and ZFC which is a qualitative signature of magnetic anisotropy, decreases with $ x $ for $ 0.125 \leq x \leq 0.200 $ and then increases for $ 0.200 \leq x \leq 0.400 $.  The H$ _{c} $ which also signifies the magnetic anisotropy, changes with $ x $ similar to the change in bifurcation. The non$ - $monotonic behavior of both H$ _{c} $ and $ \Delta M $ with $ x $ at 10 K has been interpreted as existing two competing interactions leading FM and AFM mixed phase. This mixed magnetic phase has been explained by cluster model as follows: there are non$ - $connected FM (or AFM) clusters in the matrix of AFM (or FM) phase for the lower doping concentration range ($0.125\le x < 0.200$), connected FM clusters (may be comparable size of FM and AFM clusters or domain) at $ x = 0.200 $ while in higher doping concentration range ($0.200\le x \le 0.400$) AFM clusters are in the FM matrix. This cluster model will also well explain the conducting behavior for the entire temperature range when FM clusters are connected at  $ x \geq 0.200 $. Hence, at the boundaries of FM (or AFM) clusters in the matrix of AFM (or FM) phase the translational symmetry is broken in context of spins orientation. This leads to anisotropic exchange FM$ - $AFM interaction. The anisotropy constant $K _{an} $ of IL-LSMO-$ x $ ($ 0.125 \leq x \leq 0.400 $) determined using M$ - $H $ i.e., $ $H _{c} $, and $M _{s} $ at 10 K, varies with $ x $  similar to both H$ _{c} $ and $ \Delta M $. The origin of magnetic anisotropy in IL-LSMO-$ x $ has been interpreted as FM$ - $AFM interaction around the boundary of FM and AFM regions.

\section*{ACKNOWLADGEMENTS}
\hspace{6 mm}We thank AIRF-JNU for providing facilities for PPMS. B. K. acknowledges UGC, India for financial support through fellowship.

\bibliographystyle{apsrev4-2}
\bibliography{Ref_OMA} 

\begin{thebibliography}{54}%
\makeatletter
\providecommand \@ifxundefined [1]{%
 \@ifx{#1\undefined}
}%
\providecommand \@ifnum [1]{%
 \ifnum #1\expandafter \@firstoftwo
 \else \expandafter \@secondoftwo
 \fi
}%
\providecommand \@ifx [1]{%
 \ifx #1\expandafter \@firstoftwo
 \else \expandafter \@secondoftwo
 \fi
}%
\providecommand \natexlab [1]{#1}%
\providecommand \enquote  [1]{``#1''}%
\providecommand \bibnamefont  [1]{#1}%
\providecommand \bibfnamefont [1]{#1}%
\providecommand \citenamefont [1]{#1}%
\providecommand \href@noop [0]{\@secondoftwo}%
\providecommand \href [0]{\begingroup \@sanitize@url \@href}%
\providecommand \@href[1]{\@@startlink{#1}\@@href}%
\providecommand \@@href[1]{\endgroup#1\@@endlink}%
\providecommand \@sanitize@url [0]{\catcode `\\12\catcode `\$12\catcode
  `\&12\catcode `\#12\catcode `\^12\catcode `\_12\catcode `\%12\relax}%
\providecommand \@@startlink[1]{}%
\providecommand \@@endlink[0]{}%
\providecommand \url  [0]{\begingroup\@sanitize@url \@url }%
\providecommand \@url [1]{\endgroup\@href {#1}{\urlprefix }}%
\providecommand \urlprefix  [0]{URL }%
\providecommand \Eprint [0]{\href }%
\providecommand \doibase [0]{https://doi.org/}%
\providecommand \selectlanguage [0]{\@gobble}%
\providecommand \bibinfo  [0]{\@secondoftwo}%
\providecommand \bibfield  [0]{\@secondoftwo}%
\providecommand \translation [1]{[#1]}%
\providecommand \BibitemOpen [0]{}%
\providecommand \bibitemStop [0]{}%
\providecommand \bibitemNoStop [0]{.\EOS\space}%
\providecommand \EOS [0]{\spacefactor3000\relax}%
\providecommand \BibitemShut  [1]{\csname bibitem#1\endcsname}%
\let\auto@bib@innerbib\@empty
\bibitem [{\citenamefont {Kagan}\ \emph {et~al.}(2021)\citenamefont {Kagan},
  \citenamefont {Kugel},\ and\ \citenamefont {Rakhmanov}}]{KAGAN20211}%
  \BibitemOpen
  \bibfield  {author} {\bibinfo {author} {\bibfnamefont {M.}~\bibnamefont
  {Kagan}}, \bibinfo {author} {\bibfnamefont {K.}~\bibnamefont {Kugel}},\ and\
  \bibinfo {author} {\bibfnamefont {A.}~\bibnamefont {Rakhmanov}},\ }\href
  {https://doi.org/https://doi.org/10.1016/j.physrep.2021.02.004} {\bibfield
  {journal} {\bibinfo  {journal} {Physics Reports}\ }\textbf {\bibinfo {volume}
  {916}},\ \bibinfo {pages} {1} (\bibinfo {year} {2021})},\ \bibinfo {note}
  {electronic phase separation: recent progress in the old problem}\BibitemShut
  {NoStop}%
\bibitem [{\citenamefont {Baldini}\ \emph {et~al.}(2015)\citenamefont
  {Baldini}, \citenamefont {Muramatsu}, \citenamefont {Sherafati},
  \citenamefont {Mao}, \citenamefont {Malavasi}, \citenamefont {Postorino},
  \citenamefont {Satpathy},\ and\ \citenamefont
  {Struzhkin}}]{baldini2015origin}%
  \BibitemOpen
  \bibfield  {author} {\bibinfo {author} {\bibfnamefont {M.}~\bibnamefont
  {Baldini}}, \bibinfo {author} {\bibfnamefont {T.}~\bibnamefont {Muramatsu}},
  \bibinfo {author} {\bibfnamefont {M.}~\bibnamefont {Sherafati}}, \bibinfo
  {author} {\bibfnamefont {H.-k.}\ \bibnamefont {Mao}}, \bibinfo {author}
  {\bibfnamefont {L.}~\bibnamefont {Malavasi}}, \bibinfo {author}
  {\bibfnamefont {P.}~\bibnamefont {Postorino}}, \bibinfo {author}
  {\bibfnamefont {S.}~\bibnamefont {Satpathy}},\ and\ \bibinfo {author}
  {\bibfnamefont {V.~V.}\ \bibnamefont {Struzhkin}},\ }\href@noop {} {\bibfield
   {journal} {\bibinfo  {journal} {Proceedings of the National Academy of
  Sciences}\ }\textbf {\bibinfo {volume} {112}},\ \bibinfo {pages} {10869}
  (\bibinfo {year} {2015})}\BibitemShut {NoStop}%
\bibitem [{\citenamefont {Tokura}(2006)}]{Tokura_2006}%
  \BibitemOpen
  \bibfield  {author} {\bibinfo {author} {\bibfnamefont {Y.}~\bibnamefont
  {Tokura}},\ }\href {https://doi.org/10.1088/0034-4885/69/3/r06} {\bibfield
  {journal} {\bibinfo  {journal} {Reports on Progress in Physics}\ }\textbf
  {\bibinfo {volume} {69}},\ \bibinfo {pages} {797} (\bibinfo {year}
  {2006})}\BibitemShut {NoStop}%
\bibitem [{\citenamefont {Paraskevopoulos}\ \emph {et~al.}(2000)\citenamefont
  {Paraskevopoulos}, \citenamefont {Mayr}, \citenamefont {Hartinger},
  \citenamefont {Pimenov}, \citenamefont {Hemberger}, \citenamefont
  {Lunkenheimer}, \citenamefont {Loidl}, \citenamefont {Mukhin}, \citenamefont
  {Ivanov},\ and\ \citenamefont {Balbashov}}]{PARASKEVOPOULOS2000118}%
  \BibitemOpen
  \bibfield  {author} {\bibinfo {author} {\bibfnamefont {M.}~\bibnamefont
  {Paraskevopoulos}}, \bibinfo {author} {\bibfnamefont {F.}~\bibnamefont
  {Mayr}}, \bibinfo {author} {\bibfnamefont {C.}~\bibnamefont {Hartinger}},
  \bibinfo {author} {\bibfnamefont {A.}~\bibnamefont {Pimenov}}, \bibinfo
  {author} {\bibfnamefont {J.}~\bibnamefont {Hemberger}}, \bibinfo {author}
  {\bibfnamefont {P.}~\bibnamefont {Lunkenheimer}}, \bibinfo {author}
  {\bibfnamefont {A.}~\bibnamefont {Loidl}}, \bibinfo {author} {\bibfnamefont
  {A.}~\bibnamefont {Mukhin}}, \bibinfo {author} {\bibfnamefont
  {V.}~\bibnamefont {Ivanov}},\ and\ \bibinfo {author} {\bibfnamefont
  {A.}~\bibnamefont {Balbashov}},\ }\href
  {https://doi.org/https://doi.org/10.1016/S0304-8853(99)00722-2} {\bibfield
  {journal} {\bibinfo  {journal} {Journal of Magnetism and Magnetic Materials}\
  }\textbf {\bibinfo {volume} {211}},\ \bibinfo {pages} {118} (\bibinfo {year}
  {2000})}\BibitemShut {NoStop}%
\bibitem [{\citenamefont {Ramakrishnan}\ \emph {et~al.}(2004)\citenamefont
  {Ramakrishnan}, \citenamefont {Krishnamurthy}, \citenamefont {Hassan},\ and\
  \citenamefont {Pai}}]{PhysRevLett.92.157203}%
  \BibitemOpen
  \bibfield  {author} {\bibinfo {author} {\bibfnamefont {T.~V.}\ \bibnamefont
  {Ramakrishnan}}, \bibinfo {author} {\bibfnamefont {H.~R.}\ \bibnamefont
  {Krishnamurthy}}, \bibinfo {author} {\bibfnamefont {S.~R.}\ \bibnamefont
  {Hassan}},\ and\ \bibinfo {author} {\bibfnamefont {G.~V.}\ \bibnamefont
  {Pai}},\ }\href {https://doi.org/10.1103/PhysRevLett.92.157203} {\bibfield
  {journal} {\bibinfo  {journal} {Phys. Rev. Lett.}\ }\textbf {\bibinfo
  {volume} {92}},\ \bibinfo {pages} {157203} (\bibinfo {year}
  {2004})}\BibitemShut {NoStop}%
\bibitem [{\citenamefont {Urushibara}\ \emph
  {et~al.}(1995{\natexlab{a}})\citenamefont {Urushibara}, \citenamefont
  {Moritomo}, \citenamefont {Arima}, \citenamefont {Asamitsu}, \citenamefont
  {Kido},\ and\ \citenamefont {Tokura}}]{PhysRevB.51.14103}%
  \BibitemOpen
  \bibfield  {author} {\bibinfo {author} {\bibfnamefont {A.}~\bibnamefont
  {Urushibara}}, \bibinfo {author} {\bibfnamefont {Y.}~\bibnamefont
  {Moritomo}}, \bibinfo {author} {\bibfnamefont {T.}~\bibnamefont {Arima}},
  \bibinfo {author} {\bibfnamefont {A.}~\bibnamefont {Asamitsu}}, \bibinfo
  {author} {\bibfnamefont {G.}~\bibnamefont {Kido}},\ and\ \bibinfo {author}
  {\bibfnamefont {Y.}~\bibnamefont {Tokura}},\ }\href
  {https://doi.org/10.1103/PhysRevB.51.14103} {\bibfield  {journal} {\bibinfo
  {journal} {Phys. Rev. B}\ }\textbf {\bibinfo {volume} {51}},\ \bibinfo
  {pages} {14103} (\bibinfo {year} {1995}{\natexlab{a}})}\BibitemShut {NoStop}%
\bibitem [{\citenamefont {Ni}\ \emph {et~al.}(2021)\citenamefont {Ni},
  \citenamefont {Zhao}, \citenamefont {Zhang}, \citenamefont {Hu},
  \citenamefont {Kimchi},\ and\ \citenamefont {Cao}}]{PhysRevB.103.L161105}%
  \BibitemOpen
  \bibfield  {author} {\bibinfo {author} {\bibfnamefont {Y.}~\bibnamefont
  {Ni}}, \bibinfo {author} {\bibfnamefont {H.}~\bibnamefont {Zhao}}, \bibinfo
  {author} {\bibfnamefont {Y.}~\bibnamefont {Zhang}}, \bibinfo {author}
  {\bibfnamefont {B.}~\bibnamefont {Hu}}, \bibinfo {author} {\bibfnamefont
  {I.}~\bibnamefont {Kimchi}},\ and\ \bibinfo {author} {\bibfnamefont
  {G.}~\bibnamefont {Cao}},\ }\href
  {https://doi.org/10.1103/PhysRevB.103.L161105} {\bibfield  {journal}
  {\bibinfo  {journal} {Phys. Rev. B}\ }\textbf {\bibinfo {volume} {103}},\
  \bibinfo {pages} {L161105} (\bibinfo {year} {2021})}\BibitemShut {NoStop}%
\bibitem [{\citenamefont {Beaudin}\ \emph {et~al.}(2022)\citenamefont
  {Beaudin}, \citenamefont {Fournier}, \citenamefont {Bianchi}, \citenamefont
  {Nicklas}, \citenamefont {Kenzelmann}, \citenamefont {Laver},\ and\
  \citenamefont {Witczak-Krempa}}]{PhysRevB.105.035104}%
  \BibitemOpen
  \bibfield  {author} {\bibinfo {author} {\bibfnamefont {G.}~\bibnamefont
  {Beaudin}}, \bibinfo {author} {\bibfnamefont {L.~M.}\ \bibnamefont
  {Fournier}}, \bibinfo {author} {\bibfnamefont {A.~D.}\ \bibnamefont
  {Bianchi}}, \bibinfo {author} {\bibfnamefont {M.}~\bibnamefont {Nicklas}},
  \bibinfo {author} {\bibfnamefont {M.}~\bibnamefont {Kenzelmann}}, \bibinfo
  {author} {\bibfnamefont {M.}~\bibnamefont {Laver}},\ and\ \bibinfo {author}
  {\bibfnamefont {W.}~\bibnamefont {Witczak-Krempa}},\ }\href
  {https://doi.org/10.1103/PhysRevB.105.035104} {\bibfield  {journal} {\bibinfo
   {journal} {Phys. Rev. B}\ }\textbf {\bibinfo {volume} {105}},\ \bibinfo
  {pages} {035104} (\bibinfo {year} {2022})}\BibitemShut {NoStop}%
\bibitem [{\citenamefont {Castleton}\ and\ \citenamefont
  {Altarelli}(2000)}]{PhysRevB.62.1033}%
  \BibitemOpen
  \bibfield  {author} {\bibinfo {author} {\bibfnamefont {C.~W.~M.}\
  \bibnamefont {Castleton}}\ and\ \bibinfo {author} {\bibfnamefont
  {M.}~\bibnamefont {Altarelli}},\ }\href
  {https://doi.org/10.1103/PhysRevB.62.1033} {\bibfield  {journal} {\bibinfo
  {journal} {Phys. Rev. B}\ }\textbf {\bibinfo {volume} {62}},\ \bibinfo
  {pages} {1033} (\bibinfo {year} {2000})}\BibitemShut {NoStop}%
\bibitem [{\citenamefont {Mahendiran}\ \emph {et~al.}(1995)\citenamefont
  {Mahendiran}, \citenamefont {Raychaudhuri}, \citenamefont {Chainani},
  \citenamefont {Sarma},\ and\ \citenamefont {Roy}}]{mahendiran1995}%
  \BibitemOpen
  \bibfield  {author} {\bibinfo {author} {\bibfnamefont {R.}~\bibnamefont
  {Mahendiran}}, \bibinfo {author} {\bibfnamefont {A.}~\bibnamefont
  {Raychaudhuri}}, \bibinfo {author} {\bibfnamefont {A.}~\bibnamefont
  {Chainani}}, \bibinfo {author} {\bibfnamefont {D.~D.}\ \bibnamefont
  {Sarma}},\ and\ \bibinfo {author} {\bibfnamefont {S.}~\bibnamefont {Roy}},\
  }\href@noop {} {\bibfield  {journal} {\bibinfo  {journal} {Applied physics
  letters}\ }\textbf {\bibinfo {volume} {66}},\ \bibinfo {pages} {233}
  (\bibinfo {year} {1995})}\BibitemShut {NoStop}%
\bibitem [{\citenamefont {Salazar-Muñoz}\ \emph {et~al.}(2022)\citenamefont
  {Salazar-Muñoz}, \citenamefont {{Lobo Guerrero}},\ and\ \citenamefont
  {Palomares-Sánchez}}]{SALAZARMUNOZ2022169787}%
  \BibitemOpen
  \bibfield  {author} {\bibinfo {author} {\bibfnamefont {V.}~\bibnamefont
  {Salazar-Muñoz}}, \bibinfo {author} {\bibfnamefont {A.}~\bibnamefont {{Lobo
  Guerrero}}},\ and\ \bibinfo {author} {\bibfnamefont {S.}~\bibnamefont
  {Palomares-Sánchez}},\ }\href
  {https://doi.org/https://doi.org/10.1016/j.jmmm.2022.169787} {\bibfield
  {journal} {\bibinfo  {journal} {Journal of Magnetism and Magnetic Materials}\
  }\textbf {\bibinfo {volume} {562}},\ \bibinfo {pages} {169787} (\bibinfo
  {year} {2022})}\BibitemShut {NoStop}%
\bibitem [{\citenamefont {Ram}\ \emph {et~al.}(2018)\citenamefont {Ram},
  \citenamefont {Prakash}, \citenamefont {Naresh}, \citenamefont {Kumar},
  \citenamefont {Sarmash}, \citenamefont {Subbarao}, \citenamefont {Kumar},
  \citenamefont {Kumar},\ and\ \citenamefont {Naidu}}]{Ram2018-we}%
  \BibitemOpen
  \bibfield  {author} {\bibinfo {author} {\bibfnamefont {N.~R.}\ \bibnamefont
  {Ram}}, \bibinfo {author} {\bibfnamefont {M.}~\bibnamefont {Prakash}},
  \bibinfo {author} {\bibfnamefont {U.}~\bibnamefont {Naresh}}, \bibinfo
  {author} {\bibfnamefont {N.~S.}\ \bibnamefont {Kumar}}, \bibinfo {author}
  {\bibfnamefont {T.~S.}\ \bibnamefont {Sarmash}}, \bibinfo {author}
  {\bibfnamefont {T.}~\bibnamefont {Subbarao}}, \bibinfo {author}
  {\bibfnamefont {R.~J.}\ \bibnamefont {Kumar}}, \bibinfo {author}
  {\bibfnamefont {G.~R.}\ \bibnamefont {Kumar}},\ and\ \bibinfo {author}
  {\bibfnamefont {K.~C.~B.}\ \bibnamefont {Naidu}},\ }\href@noop {} {\bibfield
  {journal} {\bibinfo  {journal} {Journal of Superconductivity and Novel
  Magnetism}\ }\textbf {\bibinfo {volume} {31}},\ \bibinfo {pages} {1971}
  (\bibinfo {year} {2018})}\BibitemShut {NoStop}%
\bibitem [{\citenamefont {Xia}\ \emph {et~al.}(2020)\citenamefont {Xia},
  \citenamefont {Pei}, \citenamefont {Leng},\ and\ \citenamefont
  {Zhu}}]{Xia2020-tx}%
  \BibitemOpen
  \bibfield  {author} {\bibinfo {author} {\bibfnamefont {W.}~\bibnamefont
  {Xia}}, \bibinfo {author} {\bibfnamefont {Z.}~\bibnamefont {Pei}}, \bibinfo
  {author} {\bibfnamefont {K.}~\bibnamefont {Leng}},\ and\ \bibinfo {author}
  {\bibfnamefont {X.}~\bibnamefont {Zhu}},\ }\href@noop {} {\bibfield
  {journal} {\bibinfo  {journal} {Nanoscale Research Letters}\ }\textbf
  {\bibinfo {volume} {15}},\ \bibinfo {pages} {9} (\bibinfo {year}
  {2020})}\BibitemShut {NoStop}%
\bibitem [{\citenamefont {Göbel}\ \emph {et~al.}(2021)\citenamefont {Göbel},
  \citenamefont {Mertig},\ and\ \citenamefont {Tretiakov}}]{GOBEL20211}%
  \BibitemOpen
  \bibfield  {author} {\bibinfo {author} {\bibfnamefont {B.}~\bibnamefont
  {Göbel}}, \bibinfo {author} {\bibfnamefont {I.}~\bibnamefont {Mertig}},\
  and\ \bibinfo {author} {\bibfnamefont {O.~A.}\ \bibnamefont {Tretiakov}},\
  }\href {https://doi.org/https://doi.org/10.1016/j.physrep.2020.10.001}
  {\bibfield  {journal} {\bibinfo  {journal} {Physics Reports}\ }\textbf
  {\bibinfo {volume} {895}},\ \bibinfo {pages} {1} (\bibinfo {year} {2021})},\
  \bibinfo {note} {beyond skyrmions: Review and perspectives of alternative
  magnetic quasiparticles}\BibitemShut {NoStop}%
\bibitem [{\citenamefont {Mohanta}\ \emph {et~al.}(2019)\citenamefont
  {Mohanta}, \citenamefont {Dagotto},\ and\ \citenamefont
  {Okamoto}}]{PhysRevB.100.064429}%
  \BibitemOpen
  \bibfield  {author} {\bibinfo {author} {\bibfnamefont {N.}~\bibnamefont
  {Mohanta}}, \bibinfo {author} {\bibfnamefont {E.}~\bibnamefont {Dagotto}},\
  and\ \bibinfo {author} {\bibfnamefont {S.}~\bibnamefont {Okamoto}},\ }\href
  {https://doi.org/10.1103/PhysRevB.100.064429} {\bibfield  {journal} {\bibinfo
   {journal} {Phys. Rev. B}\ }\textbf {\bibinfo {volume} {100}},\ \bibinfo
  {pages} {064429} (\bibinfo {year} {2019})}\BibitemShut {NoStop}%
\bibitem [{\citenamefont {Nagao}\ \emph {et~al.}(2013)\citenamefont {Nagao},
  \citenamefont {So}, \citenamefont {Yoshida}, \citenamefont {Isobe},
  \citenamefont {Hara}, \citenamefont {Ishizuka},\ and\ \citenamefont
  {Kimoto}}]{Nagao2013-on}%
  \BibitemOpen
  \bibfield  {author} {\bibinfo {author} {\bibfnamefont {M.}~\bibnamefont
  {Nagao}}, \bibinfo {author} {\bibfnamefont {Y.-G.}\ \bibnamefont {So}},
  \bibinfo {author} {\bibfnamefont {H.}~\bibnamefont {Yoshida}}, \bibinfo
  {author} {\bibfnamefont {M.}~\bibnamefont {Isobe}}, \bibinfo {author}
  {\bibfnamefont {T.}~\bibnamefont {Hara}}, \bibinfo {author} {\bibfnamefont
  {K.}~\bibnamefont {Ishizuka}},\ and\ \bibinfo {author} {\bibfnamefont
  {K.}~\bibnamefont {Kimoto}},\ }\href@noop {} {\bibfield  {journal} {\bibinfo
  {journal} {Nature Nanotechnology}\ }\textbf {\bibinfo {volume} {8}},\
  \bibinfo {pages} {325} (\bibinfo {year} {2013})}\BibitemShut {NoStop}%
\bibitem [{\citenamefont {Tiwari}\ \emph
  {et~al.}(2020{\natexlab{a}})\citenamefont {Tiwari}, \citenamefont {Chauhan},
  \citenamefont {Kumar},\ and\ \citenamefont {Ghosh}}]{Kumar.Tiwari.2020}%
  \BibitemOpen
  \bibfield  {author} {\bibinfo {author} {\bibfnamefont {J.~K.}\ \bibnamefont
  {Tiwari}}, \bibinfo {author} {\bibfnamefont {H.~C.}\ \bibnamefont {Chauhan}},
  \bibinfo {author} {\bibfnamefont {B.}~\bibnamefont {Kumar}},\ and\ \bibinfo
  {author} {\bibfnamefont {S.}~\bibnamefont {Ghosh}},\ }\href
  {https://doi.org/10.1088/1361-648X/ab6d14} {\bibfield  {journal} {\bibinfo
  {journal} {Journal of Physics: Condensed Matter}\ }\textbf {\bibinfo {volume}
  {32}},\ \bibinfo {pages} {195803} (\bibinfo {year}
  {2020}{\natexlab{a}})}\BibitemShut {NoStop}%
\bibitem [{\citenamefont {Kumar}\ \emph {et~al.}(2021)\citenamefont {Kumar},
  \citenamefont {Tiwari}, \citenamefont {Chauhan},\ and\ \citenamefont
  {Ghosh}}]{Kumar2021-sp}%
  \BibitemOpen
  \bibfield  {author} {\bibinfo {author} {\bibfnamefont {B.}~\bibnamefont
  {Kumar}}, \bibinfo {author} {\bibfnamefont {J.~K.}\ \bibnamefont {Tiwari}},
  \bibinfo {author} {\bibfnamefont {H.~C.}\ \bibnamefont {Chauhan}},\ and\
  \bibinfo {author} {\bibfnamefont {S.}~\bibnamefont {Ghosh}},\ }\href@noop {}
  {\bibfield  {journal} {\bibinfo  {journal} {Scientific Reports}\ }\textbf
  {\bibinfo {volume} {11}},\ \bibinfo {pages} {21184} (\bibinfo {year}
  {2021})}\BibitemShut {NoStop}%
\bibitem [{\citenamefont {Tiwari}\ \emph
  {et~al.}(2021{\natexlab{a}})\citenamefont {Tiwari}, \citenamefont {Kumar},
  \citenamefont {Chauhan},\ and\ \citenamefont {Ghosh}}]{TIWARI2021168020}%
  \BibitemOpen
  \bibfield  {author} {\bibinfo {author} {\bibfnamefont {J.~K.}\ \bibnamefont
  {Tiwari}}, \bibinfo {author} {\bibfnamefont {B.}~\bibnamefont {Kumar}},
  \bibinfo {author} {\bibfnamefont {H.~C.}\ \bibnamefont {Chauhan}},\ and\
  \bibinfo {author} {\bibfnamefont {S.}~\bibnamefont {Ghosh}},\ }\href
  {https://doi.org/https://doi.org/10.1016/j.jmmm.2021.168020} {\bibfield
  {journal} {\bibinfo  {journal} {Journal of Magnetism and Magnetic Materials}\
  }\textbf {\bibinfo {volume} {535}},\ \bibinfo {pages} {168020} (\bibinfo
  {year} {2021}{\natexlab{a}})}\BibitemShut {NoStop}%
\bibitem [{\citenamefont {Tiwari}\ \emph
  {et~al.}(2021{\natexlab{b}})\citenamefont {Tiwari}, \citenamefont {Kumar},
  \citenamefont {Chauhan},\ and\ \citenamefont {Ghosh}}]{Tiwari2021-su}%
  \BibitemOpen
  \bibfield  {author} {\bibinfo {author} {\bibfnamefont {J.~K.}\ \bibnamefont
  {Tiwari}}, \bibinfo {author} {\bibfnamefont {B.}~\bibnamefont {Kumar}},
  \bibinfo {author} {\bibfnamefont {H.~C.}\ \bibnamefont {Chauhan}},\ and\
  \bibinfo {author} {\bibfnamefont {S.}~\bibnamefont {Ghosh}},\ }\href@noop {}
  {\bibfield  {journal} {\bibinfo  {journal} {Scientific Reports}\ }\textbf
  {\bibinfo {volume} {11}},\ \bibinfo {pages} {14117} (\bibinfo {year}
  {2021}{\natexlab{b}})}\BibitemShut {NoStop}%
\bibitem [{\citenamefont {Wang}\ \emph {et~al.}(2022)\citenamefont {Wang},
  \citenamefont {Wang}, \citenamefont {Zhou}, \citenamefont {Wang},
  \citenamefont {Shi},\ and\ \citenamefont {Du}}]{Wang2022-ku}%
  \BibitemOpen
  \bibfield  {author} {\bibinfo {author} {\bibfnamefont {C.-J.}\ \bibnamefont
  {Wang}}, \bibinfo {author} {\bibfnamefont {P.}~\bibnamefont {Wang}}, \bibinfo
  {author} {\bibfnamefont {Y.}~\bibnamefont {Zhou}}, \bibinfo {author}
  {\bibfnamefont {W.}~\bibnamefont {Wang}}, \bibinfo {author} {\bibfnamefont
  {F.}~\bibnamefont {Shi}},\ and\ \bibinfo {author} {\bibfnamefont
  {J.}~\bibnamefont {Du}},\ }\href@noop {} {\bibfield  {journal} {\bibinfo
  {journal} {npj Quantum Materials}\ }\textbf {\bibinfo {volume} {7}},\
  \bibinfo {pages} {78} (\bibinfo {year} {2022})}\BibitemShut {NoStop}%
\bibitem [{\citenamefont {Yu}\ \emph {et~al.}(2014)\citenamefont {Yu},
  \citenamefont {Tokunaga}, \citenamefont {Kaneko}, \citenamefont {Zhang},
  \citenamefont {Kimoto}, \citenamefont {Matsui}, \citenamefont {Taguchi},\
  and\ \citenamefont {Tokura}}]{Yu2014-if}%
  \BibitemOpen
  \bibfield  {author} {\bibinfo {author} {\bibfnamefont {X.~Z.}\ \bibnamefont
  {Yu}}, \bibinfo {author} {\bibfnamefont {Y.}~\bibnamefont {Tokunaga}},
  \bibinfo {author} {\bibfnamefont {Y.}~\bibnamefont {Kaneko}}, \bibinfo
  {author} {\bibfnamefont {W.~Z.}\ \bibnamefont {Zhang}}, \bibinfo {author}
  {\bibfnamefont {K.}~\bibnamefont {Kimoto}}, \bibinfo {author} {\bibfnamefont
  {Y.}~\bibnamefont {Matsui}}, \bibinfo {author} {\bibfnamefont
  {Y.}~\bibnamefont {Taguchi}},\ and\ \bibinfo {author} {\bibfnamefont
  {Y.}~\bibnamefont {Tokura}},\ }\href@noop {} {\bibfield  {journal} {\bibinfo
  {journal} {Nature Communications}\ }\textbf {\bibinfo {volume} {5}},\
  \bibinfo {pages} {3198} (\bibinfo {year} {2014})}\BibitemShut {NoStop}%
\bibitem [{\citenamefont {G{\"o}bel}\ \emph {et~al.}(2019)\citenamefont
  {G{\"o}bel}, \citenamefont {Henk},\ and\ \citenamefont
  {Mertig}}]{Gobel2019-mu}%
  \BibitemOpen
  \bibfield  {author} {\bibinfo {author} {\bibfnamefont {B.}~\bibnamefont
  {G{\"o}bel}}, \bibinfo {author} {\bibfnamefont {J.}~\bibnamefont {Henk}},\
  and\ \bibinfo {author} {\bibfnamefont {I.}~\bibnamefont {Mertig}},\
  }\href@noop {} {\bibfield  {journal} {\bibinfo  {journal} {Scientific
  Reports}\ }\textbf {\bibinfo {volume} {9}},\ \bibinfo {pages} {9521}
  (\bibinfo {year} {2019})}\BibitemShut {NoStop}%
\bibitem [{\citenamefont {Ruddlesden}\ and\ \citenamefont
  {Popper}(1958)}]{Ruddlesden.1958}%
  \BibitemOpen
  \bibfield  {author} {\bibinfo {author} {\bibfnamefont {S.~N.}\ \bibnamefont
  {Ruddlesden}}\ and\ \bibinfo {author} {\bibfnamefont {P.}~\bibnamefont
  {Popper}},\ }\href
  {https://doi.org/https://doi.org/10.1107/S0365110X58000128} {\bibfield
  {journal} {\bibinfo  {journal} {Acta Crystallographica}\ }\textbf {\bibinfo
  {volume} {11}},\ \bibinfo {pages} {54} (\bibinfo {year} {1958})}\BibitemShut
  {NoStop}%
\bibitem [{\citenamefont {von Helmolt}\ \emph {et~al.}(1993)\citenamefont {von
  Helmolt}, \citenamefont {Wecker}, \citenamefont {Holzapfel}, \citenamefont
  {Schultz},\ and\ \citenamefont {Samwer}}]{PhysRevLett.71.2331}%
  \BibitemOpen
  \bibfield  {author} {\bibinfo {author} {\bibfnamefont {R.}~\bibnamefont {von
  Helmolt}}, \bibinfo {author} {\bibfnamefont {J.}~\bibnamefont {Wecker}},
  \bibinfo {author} {\bibfnamefont {B.}~\bibnamefont {Holzapfel}}, \bibinfo
  {author} {\bibfnamefont {L.}~\bibnamefont {Schultz}},\ and\ \bibinfo {author}
  {\bibfnamefont {K.}~\bibnamefont {Samwer}},\ }\href
  {https://doi.org/10.1103/PhysRevLett.71.2331} {\bibfield  {journal} {\bibinfo
   {journal} {Phys. Rev. Lett.}\ }\textbf {\bibinfo {volume} {71}},\ \bibinfo
  {pages} {2331} (\bibinfo {year} {1993})}\BibitemShut {NoStop}%
\bibitem [{\citenamefont {Markovich}\ \emph {et~al.}(2014)\citenamefont
  {Markovich}, \citenamefont {Wisniewski},\ and\ \citenamefont
  {Szymczak}}]{MARKOVICH20141}%
  \BibitemOpen
  \bibfield  {author} {\bibinfo {author} {\bibfnamefont {V.}~\bibnamefont
  {Markovich}}, \bibinfo {author} {\bibfnamefont {A.}~\bibnamefont
  {Wisniewski}},\ and\ \bibinfo {author} {\bibfnamefont {H.}~\bibnamefont
  {Szymczak}}\ }(\bibinfo  {publisher} {Elsevier},\ \bibinfo {year} {2014})\
  pp.\ \bibinfo {pages} {1--201}\BibitemShut {NoStop}%
\bibitem [{\citenamefont {Moussa}\ \emph {et~al.}(1996)\citenamefont {Moussa},
  \citenamefont {Hennion}, \citenamefont {Rodriguez-Carvajal}, \citenamefont
  {Moudden}, \citenamefont {Pinsard},\ and\ \citenamefont
  {Revcolevschi}}]{PhysRevB.54.15149}%
  \BibitemOpen
  \bibfield  {author} {\bibinfo {author} {\bibfnamefont {F.}~\bibnamefont
  {Moussa}}, \bibinfo {author} {\bibfnamefont {M.}~\bibnamefont {Hennion}},
  \bibinfo {author} {\bibfnamefont {J.}~\bibnamefont {Rodriguez-Carvajal}},
  \bibinfo {author} {\bibfnamefont {H.}~\bibnamefont {Moudden}}, \bibinfo
  {author} {\bibfnamefont {L.}~\bibnamefont {Pinsard}},\ and\ \bibinfo {author}
  {\bibfnamefont {A.}~\bibnamefont {Revcolevschi}},\ }\href
  {https://doi.org/10.1103/PhysRevB.54.15149} {\bibfield  {journal} {\bibinfo
  {journal} {Phys. Rev. B}\ }\textbf {\bibinfo {volume} {54}},\ \bibinfo
  {pages} {15149} (\bibinfo {year} {1996})}\BibitemShut {NoStop}%
\bibitem [{\citenamefont {de~Gennes}(1960)}]{PhysRev.118.141}%
  \BibitemOpen
  \bibfield  {author} {\bibinfo {author} {\bibfnamefont {P.~G.}\ \bibnamefont
  {de~Gennes}},\ }\href {https://doi.org/10.1103/PhysRev.118.141} {\bibfield
  {journal} {\bibinfo  {journal} {Phys. Rev.}\ }\textbf {\bibinfo {volume}
  {118}},\ \bibinfo {pages} {141} (\bibinfo {year} {1960})}\BibitemShut
  {NoStop}%
\bibitem [{\citenamefont {Dabrowski}\ \emph
  {et~al.}(1999{\natexlab{a}})\citenamefont {Dabrowski}, \citenamefont {Xiong},
  \citenamefont {Bukowski}, \citenamefont {Dybzinski}, \citenamefont {Klamut},
  \citenamefont {Siewenie}, \citenamefont {Chmaissem}, \citenamefont {Shaffer},
  \citenamefont {Kimball}, \citenamefont {Jorgensen},\ and\ \citenamefont
  {Short}}]{PhysRevB.60.7006}%
  \BibitemOpen
  \bibfield  {author} {\bibinfo {author} {\bibfnamefont {B.}~\bibnamefont
  {Dabrowski}}, \bibinfo {author} {\bibfnamefont {X.}~\bibnamefont {Xiong}},
  \bibinfo {author} {\bibfnamefont {Z.}~\bibnamefont {Bukowski}}, \bibinfo
  {author} {\bibfnamefont {R.}~\bibnamefont {Dybzinski}}, \bibinfo {author}
  {\bibfnamefont {P.~W.}\ \bibnamefont {Klamut}}, \bibinfo {author}
  {\bibfnamefont {J.~E.}\ \bibnamefont {Siewenie}}, \bibinfo {author}
  {\bibfnamefont {O.}~\bibnamefont {Chmaissem}}, \bibinfo {author}
  {\bibfnamefont {J.}~\bibnamefont {Shaffer}}, \bibinfo {author} {\bibfnamefont
  {C.~W.}\ \bibnamefont {Kimball}}, \bibinfo {author} {\bibfnamefont {J.~D.}\
  \bibnamefont {Jorgensen}},\ and\ \bibinfo {author} {\bibfnamefont
  {S.}~\bibnamefont {Short}},\ }\href
  {https://doi.org/10.1103/PhysRevB.60.7006} {\bibfield  {journal} {\bibinfo
  {journal} {Phys. Rev. B}\ }\textbf {\bibinfo {volume} {60}},\ \bibinfo
  {pages} {7006} (\bibinfo {year} {1999}{\natexlab{a}})}\BibitemShut {NoStop}%
\bibitem [{\citenamefont {Urushibara}\ \emph
  {et~al.}(1995{\natexlab{b}})\citenamefont {Urushibara}, \citenamefont
  {Moritomo}, \citenamefont {Arima}, \citenamefont {Asamitsu}, \citenamefont
  {Kido},\ and\ \citenamefont {Tokura}}]{urushibara1995insulator}%
  \BibitemOpen
  \bibfield  {author} {\bibinfo {author} {\bibfnamefont {A.}~\bibnamefont
  {Urushibara}}, \bibinfo {author} {\bibfnamefont {Y.}~\bibnamefont
  {Moritomo}}, \bibinfo {author} {\bibfnamefont {T.}~\bibnamefont {Arima}},
  \bibinfo {author} {\bibfnamefont {A.}~\bibnamefont {Asamitsu}}, \bibinfo
  {author} {\bibfnamefont {G.}~\bibnamefont {Kido}},\ and\ \bibinfo {author}
  {\bibfnamefont {Y.}~\bibnamefont {Tokura}},\ }\href@noop {} {\bibfield
  {journal} {\bibinfo  {journal} {Physical Review B}\ }\textbf {\bibinfo
  {volume} {51}},\ \bibinfo {pages} {14103} (\bibinfo {year}
  {1995}{\natexlab{b}})}\BibitemShut {NoStop}%
\bibitem [{\citenamefont {Kawano}\ \emph {et~al.}(1997)\citenamefont {Kawano},
  \citenamefont {Kajimoto}, \citenamefont {Yoshizawa}, \citenamefont {Tomioka},
  \citenamefont {Kuwahara},\ and\ \citenamefont
  {Tokura}}]{PhysRevLett.78.4253}%
  \BibitemOpen
  \bibfield  {author} {\bibinfo {author} {\bibfnamefont {H.}~\bibnamefont
  {Kawano}}, \bibinfo {author} {\bibfnamefont {R.}~\bibnamefont {Kajimoto}},
  \bibinfo {author} {\bibfnamefont {H.}~\bibnamefont {Yoshizawa}}, \bibinfo
  {author} {\bibfnamefont {Y.}~\bibnamefont {Tomioka}}, \bibinfo {author}
  {\bibfnamefont {H.}~\bibnamefont {Kuwahara}},\ and\ \bibinfo {author}
  {\bibfnamefont {Y.}~\bibnamefont {Tokura}},\ }\href
  {https://doi.org/10.1103/PhysRevLett.78.4253} {\bibfield  {journal} {\bibinfo
   {journal} {Phys. Rev. Lett.}\ }\textbf {\bibinfo {volume} {78}},\ \bibinfo
  {pages} {4253} (\bibinfo {year} {1997})}\BibitemShut {NoStop}%
\bibitem [{\citenamefont {Kumar}\ \emph {et~al.}(1998)\citenamefont {Kumar},
  \citenamefont {Joy},\ and\ \citenamefont {Date}}]{kumar1998origin}%
  \BibitemOpen
  \bibfield  {author} {\bibinfo {author} {\bibfnamefont {P.~A.}\ \bibnamefont
  {Kumar}}, \bibinfo {author} {\bibfnamefont {P.}~\bibnamefont {Joy}},\ and\
  \bibinfo {author} {\bibfnamefont {S.}~\bibnamefont {Date}},\ }\href@noop {}
  {\bibfield  {journal} {\bibinfo  {journal} {Journal of Physics: Condensed
  Matter}\ }\textbf {\bibinfo {volume} {10}},\ \bibinfo {pages} {L487}
  (\bibinfo {year} {1998})}\BibitemShut {NoStop}%
\bibitem [{\citenamefont {Joy}\ \emph {et~al.}(1998)\citenamefont {Joy},
  \citenamefont {Kumar},\ and\ \citenamefont {Date}}]{joy1998}%
  \BibitemOpen
  \bibfield  {author} {\bibinfo {author} {\bibfnamefont {P.}~\bibnamefont
  {Joy}}, \bibinfo {author} {\bibfnamefont {P.~A.}\ \bibnamefont {Kumar}},\
  and\ \bibinfo {author} {\bibfnamefont {S.}~\bibnamefont {Date}},\ }\href@noop
  {} {\bibfield  {journal} {\bibinfo  {journal} {Journal of physics: condensed
  matter}\ }\textbf {\bibinfo {volume} {10}},\ \bibinfo {pages} {11049}
  (\bibinfo {year} {1998})}\BibitemShut {NoStop}%
\bibitem [{\citenamefont {M\"{u}hlbauer}\ \emph {et~al.}(2009)\citenamefont
  {M\"{u}hlbauer}, \citenamefont {Binz}, \citenamefont {Jonietz}, \citenamefont
  {Pfleiderer}, \citenamefont {Rosch}, \citenamefont {Neubauer}, \citenamefont
  {Georgii},\ and\ \citenamefont {Boni}}]{muhlbauer2009}%
  \BibitemOpen
  \bibfield  {author} {\bibinfo {author} {\bibfnamefont {S.}~\bibnamefont
  {M\"{u}hlbauer}}, \bibinfo {author} {\bibfnamefont {B.}~\bibnamefont {Binz}},
  \bibinfo {author} {\bibfnamefont {F.}~\bibnamefont {Jonietz}}, \bibinfo
  {author} {\bibfnamefont {C.}~\bibnamefont {Pfleiderer}}, \bibinfo {author}
  {\bibfnamefont {A.}~\bibnamefont {Rosch}}, \bibinfo {author} {\bibfnamefont
  {A.}~\bibnamefont {Neubauer}}, \bibinfo {author} {\bibfnamefont
  {R.}~\bibnamefont {Georgii}},\ and\ \bibinfo {author} {\bibfnamefont
  {P.}~\bibnamefont {Boni}},\ }\href@noop {} {\bibfield  {journal} {\bibinfo
  {journal} {Science}\ }\textbf {\bibinfo {volume} {323}},\ \bibinfo {pages}
  {915} (\bibinfo {year} {2009})}\BibitemShut {NoStop}%
\bibitem [{\citenamefont {Tonomura}\ \emph {et~al.}(2012)\citenamefont
  {Tonomura}, \citenamefont {Yu}, \citenamefont {Yanagisawa}, \citenamefont
  {Matsuda}, \citenamefont {Onose}, \citenamefont {Kanazawa}, \citenamefont
  {Park},\ and\ \citenamefont {Tokura}}]{tonomura2012}%
  \BibitemOpen
  \bibfield  {author} {\bibinfo {author} {\bibfnamefont {A.}~\bibnamefont
  {Tonomura}}, \bibinfo {author} {\bibfnamefont {X.}~\bibnamefont {Yu}},
  \bibinfo {author} {\bibfnamefont {K.}~\bibnamefont {Yanagisawa}}, \bibinfo
  {author} {\bibfnamefont {T.}~\bibnamefont {Matsuda}}, \bibinfo {author}
  {\bibfnamefont {Y.}~\bibnamefont {Onose}}, \bibinfo {author} {\bibfnamefont
  {N.}~\bibnamefont {Kanazawa}}, \bibinfo {author} {\bibfnamefont {H.~S.}\
  \bibnamefont {Park}},\ and\ \bibinfo {author} {\bibfnamefont
  {Y.}~\bibnamefont {Tokura}},\ }\href@noop {} {\bibfield  {journal} {\bibinfo
  {journal} {Nano letters}\ }\textbf {\bibinfo {volume} {12}},\ \bibinfo
  {pages} {1673} (\bibinfo {year} {2012})}\BibitemShut {NoStop}%
\bibitem [{\citenamefont {Wilhelm}\ \emph {et~al.}(2011)\citenamefont
  {Wilhelm}, \citenamefont {Baenitz}, \citenamefont {Schmidt}, \citenamefont
  {R\"o\ss{}ler}, \citenamefont {Leonov},\ and\ \citenamefont
  {Bogdanov}}]{PhysRevLett.107.127203}%
  \BibitemOpen
  \bibfield  {author} {\bibinfo {author} {\bibfnamefont {H.}~\bibnamefont
  {Wilhelm}}, \bibinfo {author} {\bibfnamefont {M.}~\bibnamefont {Baenitz}},
  \bibinfo {author} {\bibfnamefont {M.}~\bibnamefont {Schmidt}}, \bibinfo
  {author} {\bibfnamefont {U.~K.}\ \bibnamefont {R\"o\ss{}ler}}, \bibinfo
  {author} {\bibfnamefont {A.~A.}\ \bibnamefont {Leonov}},\ and\ \bibinfo
  {author} {\bibfnamefont {A.~N.}\ \bibnamefont {Bogdanov}},\ }\href
  {https://doi.org/10.1103/PhysRevLett.107.127203} {\bibfield  {journal}
  {\bibinfo  {journal} {Phys. Rev. Lett.}\ }\textbf {\bibinfo {volume} {107}},\
  \bibinfo {pages} {127203} (\bibinfo {year} {2011})}\BibitemShut {NoStop}%
\bibitem [{\citenamefont {Yu}\ \emph {et~al.}(2011)\citenamefont {Yu},
  \citenamefont {Kanazawa}, \citenamefont {Onose}, \citenamefont {Kimoto},
  \citenamefont {Zhang}, \citenamefont {Ishiwata}, \citenamefont {Matsui},\
  and\ \citenamefont {Tokura}}]{Yu2011}%
  \BibitemOpen
  \bibfield  {author} {\bibinfo {author} {\bibfnamefont {X.~Z.}\ \bibnamefont
  {Yu}}, \bibinfo {author} {\bibfnamefont {N.}~\bibnamefont {Kanazawa}},
  \bibinfo {author} {\bibfnamefont {Y.}~\bibnamefont {Onose}}, \bibinfo
  {author} {\bibfnamefont {K.}~\bibnamefont {Kimoto}}, \bibinfo {author}
  {\bibfnamefont {W.~Z.}\ \bibnamefont {Zhang}}, \bibinfo {author}
  {\bibfnamefont {S.}~\bibnamefont {Ishiwata}}, \bibinfo {author}
  {\bibfnamefont {Y.}~\bibnamefont {Matsui}},\ and\ \bibinfo {author}
  {\bibfnamefont {Y.}~\bibnamefont {Tokura}},\ }\href
  {https://doi.org/10.1038/nmat2916} {\bibfield  {journal} {\bibinfo  {journal}
  {Nature Materials}\ }\textbf {\bibinfo {volume} {10}},\ \bibinfo {pages}
  {106} (\bibinfo {year} {2011})}\BibitemShut {NoStop}%
\bibitem [{\citenamefont {Moskvin}\ \emph {et~al.}(2013)\citenamefont
  {Moskvin}, \citenamefont {Grigoriev}, \citenamefont {Dyadkin}, \citenamefont
  {Eckerlebe}, \citenamefont {Baenitz}, \citenamefont {Schmidt},\ and\
  \citenamefont {Wilhelm}}]{PhysRevLett.110.077207}%
  \BibitemOpen
  \bibfield  {author} {\bibinfo {author} {\bibfnamefont {E.}~\bibnamefont
  {Moskvin}}, \bibinfo {author} {\bibfnamefont {S.}~\bibnamefont {Grigoriev}},
  \bibinfo {author} {\bibfnamefont {V.}~\bibnamefont {Dyadkin}}, \bibinfo
  {author} {\bibfnamefont {H.}~\bibnamefont {Eckerlebe}}, \bibinfo {author}
  {\bibfnamefont {M.}~\bibnamefont {Baenitz}}, \bibinfo {author} {\bibfnamefont
  {M.}~\bibnamefont {Schmidt}},\ and\ \bibinfo {author} {\bibfnamefont
  {H.}~\bibnamefont {Wilhelm}},\ }\href
  {https://doi.org/10.1103/PhysRevLett.110.077207} {\bibfield  {journal}
  {\bibinfo  {journal} {Phys. Rev. Lett.}\ }\textbf {\bibinfo {volume} {110}},\
  \bibinfo {pages} {077207} (\bibinfo {year} {2013})}\BibitemShut {NoStop}%
\bibitem [{\citenamefont {Seki}\ \emph
  {et~al.}(2012{\natexlab{a}})\citenamefont {Seki}, \citenamefont {Yu},
  \citenamefont {Ishiwata},\ and\ \citenamefont {Tokura}}]{seki2012}%
  \BibitemOpen
  \bibfield  {author} {\bibinfo {author} {\bibfnamefont {S.}~\bibnamefont
  {Seki}}, \bibinfo {author} {\bibfnamefont {X.}~\bibnamefont {Yu}}, \bibinfo
  {author} {\bibfnamefont {S.}~\bibnamefont {Ishiwata}},\ and\ \bibinfo
  {author} {\bibfnamefont {Y.}~\bibnamefont {Tokura}},\ }\href@noop {}
  {\bibfield  {journal} {\bibinfo  {journal} {Science}\ }\textbf {\bibinfo
  {volume} {336}},\ \bibinfo {pages} {198} (\bibinfo {year}
  {2012}{\natexlab{a}})}\BibitemShut {NoStop}%
\bibitem [{\citenamefont {Seki}\ \emph
  {et~al.}(2012{\natexlab{b}})\citenamefont {Seki}, \citenamefont {Kim},
  \citenamefont {Inosov}, \citenamefont {Georgii}, \citenamefont {Keimer},
  \citenamefont {Ishiwata},\ and\ \citenamefont {Tokura}}]{PhysRevB.85.220406}%
  \BibitemOpen
  \bibfield  {author} {\bibinfo {author} {\bibfnamefont {S.}~\bibnamefont
  {Seki}}, \bibinfo {author} {\bibfnamefont {J.-H.}\ \bibnamefont {Kim}},
  \bibinfo {author} {\bibfnamefont {D.~S.}\ \bibnamefont {Inosov}}, \bibinfo
  {author} {\bibfnamefont {R.}~\bibnamefont {Georgii}}, \bibinfo {author}
  {\bibfnamefont {B.}~\bibnamefont {Keimer}}, \bibinfo {author} {\bibfnamefont
  {S.}~\bibnamefont {Ishiwata}},\ and\ \bibinfo {author} {\bibfnamefont
  {Y.}~\bibnamefont {Tokura}},\ }\href
  {https://doi.org/10.1103/PhysRevB.85.220406} {\bibfield  {journal} {\bibinfo
  {journal} {Phys. Rev. B}\ }\textbf {\bibinfo {volume} {85}},\ \bibinfo
  {pages} {220406} (\bibinfo {year} {2012}{\natexlab{b}})}\BibitemShut
  {NoStop}%
\bibitem [{\citenamefont {Lancaster}\ \emph {et~al.}(2015)\citenamefont
  {Lancaster}, \citenamefont {Williams}, \citenamefont {Thomas}, \citenamefont
  {Xiao}, \citenamefont {Pratt}, \citenamefont {Blundell}, \citenamefont
  {Loudon}, \citenamefont {Hesjedal}, \citenamefont {Clark}, \citenamefont
  {Hatton}, \citenamefont {Ciomaga~Hatnean}, \citenamefont {Keeble},\ and\
  \citenamefont {Balakrishnan}}]{PhysRevB.91.224408}%
  \BibitemOpen
  \bibfield  {author} {\bibinfo {author} {\bibfnamefont {T.}~\bibnamefont
  {Lancaster}}, \bibinfo {author} {\bibfnamefont {R.~C.}\ \bibnamefont
  {Williams}}, \bibinfo {author} {\bibfnamefont {I.~O.}\ \bibnamefont
  {Thomas}}, \bibinfo {author} {\bibfnamefont {F.}~\bibnamefont {Xiao}},
  \bibinfo {author} {\bibfnamefont {F.~L.}\ \bibnamefont {Pratt}}, \bibinfo
  {author} {\bibfnamefont {S.~J.}\ \bibnamefont {Blundell}}, \bibinfo {author}
  {\bibfnamefont {J.~C.}\ \bibnamefont {Loudon}}, \bibinfo {author}
  {\bibfnamefont {T.}~\bibnamefont {Hesjedal}}, \bibinfo {author}
  {\bibfnamefont {S.~J.}\ \bibnamefont {Clark}}, \bibinfo {author}
  {\bibfnamefont {P.~D.}\ \bibnamefont {Hatton}}, \bibinfo {author}
  {\bibfnamefont {M.}~\bibnamefont {Ciomaga~Hatnean}}, \bibinfo {author}
  {\bibfnamefont {D.~S.}\ \bibnamefont {Keeble}},\ and\ \bibinfo {author}
  {\bibfnamefont {G.}~\bibnamefont {Balakrishnan}},\ }\href
  {https://doi.org/10.1103/PhysRevB.91.224408} {\bibfield  {journal} {\bibinfo
  {journal} {Phys. Rev. B}\ }\textbf {\bibinfo {volume} {91}},\ \bibinfo
  {pages} {224408} (\bibinfo {year} {2015})}\BibitemShut {NoStop}%
\bibitem [{\citenamefont {Chauhan}\ \emph {et~al.}(2019)\citenamefont
  {Chauhan}, \citenamefont {Kumar}, \citenamefont {Tiwari},\ and\ \citenamefont
  {Ghosh}}]{PhysRevB.100.165143}%
  \BibitemOpen
  \bibfield  {author} {\bibinfo {author} {\bibfnamefont {H.~C.}\ \bibnamefont
  {Chauhan}}, \bibinfo {author} {\bibfnamefont {B.}~\bibnamefont {Kumar}},
  \bibinfo {author} {\bibfnamefont {J.~K.}\ \bibnamefont {Tiwari}},\ and\
  \bibinfo {author} {\bibfnamefont {S.}~\bibnamefont {Ghosh}},\ }\href
  {https://doi.org/10.1103/PhysRevB.100.165143} {\bibfield  {journal} {\bibinfo
   {journal} {Phys. Rev. B}\ }\textbf {\bibinfo {volume} {100}},\ \bibinfo
  {pages} {165143} (\bibinfo {year} {2019})}\BibitemShut {NoStop}%
\bibitem [{\citenamefont {Chauhan}\ \emph {et~al.}(2022)\citenamefont
  {Chauhan}, \citenamefont {Kumar}, \citenamefont {Tiwari}, \citenamefont
  {Tiwari},\ and\ \citenamefont {Ghosh}}]{PhysRevLett.128.015703}%
  \BibitemOpen
  \bibfield  {author} {\bibinfo {author} {\bibfnamefont {H.~C.}\ \bibnamefont
  {Chauhan}}, \bibinfo {author} {\bibfnamefont {B.}~\bibnamefont {Kumar}},
  \bibinfo {author} {\bibfnamefont {A.}~\bibnamefont {Tiwari}}, \bibinfo
  {author} {\bibfnamefont {J.~K.}\ \bibnamefont {Tiwari}},\ and\ \bibinfo
  {author} {\bibfnamefont {S.}~\bibnamefont {Ghosh}},\ }\href
  {https://doi.org/10.1103/PhysRevLett.128.015703} {\bibfield  {journal}
  {\bibinfo  {journal} {Phys. Rev. Lett.}\ }\textbf {\bibinfo {volume} {128}},\
  \bibinfo {pages} {015703} (\bibinfo {year} {2022})}\BibitemShut {NoStop}%
\bibitem [{\citenamefont {Tiwari}\ \emph
  {et~al.}(2020{\natexlab{b}})\citenamefont {Tiwari}, \citenamefont {Chauhan},
  \citenamefont {Kumar},\ and\ \citenamefont {Ghosh}}]{KumarTiwari2020}%
  \BibitemOpen
  \bibfield  {author} {\bibinfo {author} {\bibfnamefont {J.~K.}\ \bibnamefont
  {Tiwari}}, \bibinfo {author} {\bibfnamefont {H.~C.}\ \bibnamefont {Chauhan}},
  \bibinfo {author} {\bibfnamefont {B.}~\bibnamefont {Kumar}},\ and\ \bibinfo
  {author} {\bibfnamefont {S.}~\bibnamefont {Ghosh}},\ }\href
  {https://doi.org/10.1088/1361-648x/ab6d14} {\bibfield  {journal} {\bibinfo
  {journal} {Journal of Physics: Condensed Matter}\ }\textbf {\bibinfo {volume}
  {32}},\ \bibinfo {pages} {195803} (\bibinfo {year}
  {2020}{\natexlab{b}})}\BibitemShut {NoStop}%
\bibitem [{\citenamefont {Materials}()}]{SM}%
  \BibitemOpen
  \bibfield  {author} {\bibinfo {author} {\bibfnamefont {S.}~\bibnamefont
  {Materials}},\ }\href@noop {} {\ }\BibitemShut {NoStop}%
\bibitem [{\citenamefont {Dabrowski}\ \emph
  {et~al.}(1999{\natexlab{b}})\citenamefont {Dabrowski}, \citenamefont {Xiong},
  \citenamefont {Bukowski}, \citenamefont {Dybzinski}, \citenamefont {Klamut},
  \citenamefont {Siewenie}, \citenamefont {Chmaissem}, \citenamefont {Shaffer},
  \citenamefont {Kimball}, \citenamefont {Jorgensen},\ and\ \citenamefont
  {Short}}]{Dabrowski1999StructurepropertiesPD}%
  \BibitemOpen
  \bibfield  {author} {\bibinfo {author} {\bibfnamefont {B.}~\bibnamefont
  {Dabrowski}}, \bibinfo {author} {\bibfnamefont {X.}~\bibnamefont {Xiong}},
  \bibinfo {author} {\bibfnamefont {Z.}~\bibnamefont {Bukowski}}, \bibinfo
  {author} {\bibfnamefont {R.}~\bibnamefont {Dybzinski}}, \bibinfo {author}
  {\bibfnamefont {P.~W.}\ \bibnamefont {Klamut}}, \bibinfo {author}
  {\bibfnamefont {J.~E.}\ \bibnamefont {Siewenie}}, \bibinfo {author}
  {\bibfnamefont {O.}~\bibnamefont {Chmaissem}}, \bibinfo {author}
  {\bibfnamefont {J.}~\bibnamefont {Shaffer}}, \bibinfo {author} {\bibfnamefont
  {C.~W.}\ \bibnamefont {Kimball}}, \bibinfo {author} {\bibfnamefont {J.~D.}\
  \bibnamefont {Jorgensen}},\ and\ \bibinfo {author} {\bibfnamefont
  {S.}~\bibnamefont {Short}},\ }\href
  {https://doi.org/10.1103/PhysRevB.60.7006} {\bibfield  {journal} {\bibinfo
  {journal} {Phys. Rev. B}\ }\textbf {\bibinfo {volume} {60}},\ \bibinfo
  {pages} {7006} (\bibinfo {year} {1999}{\natexlab{b}})}\BibitemShut {NoStop}%
\bibitem [{\citenamefont {Mitra}\ \emph {et~al.}(2003)\citenamefont {Mitra},
  \citenamefont {Raychaudhuri}, \citenamefont {Mukovskii},\ and\ \citenamefont
  {Shulyatev}}]{PhysRevB.68.134428}%
  \BibitemOpen
  \bibfield  {author} {\bibinfo {author} {\bibfnamefont {J.}~\bibnamefont
  {Mitra}}, \bibinfo {author} {\bibfnamefont {A.~K.}\ \bibnamefont
  {Raychaudhuri}}, \bibinfo {author} {\bibfnamefont {Y.~M.}\ \bibnamefont
  {Mukovskii}},\ and\ \bibinfo {author} {\bibfnamefont {D.}~\bibnamefont
  {Shulyatev}},\ }\href {https://doi.org/10.1103/PhysRevB.68.134428} {\bibfield
   {journal} {\bibinfo  {journal} {Phys. Rev. B}\ }\textbf {\bibinfo {volume}
  {68}},\ \bibinfo {pages} {134428} (\bibinfo {year} {2003})}\BibitemShut
  {NoStop}%
\bibitem [{\citenamefont {Burgy}\ \emph {et~al.}(2001)\citenamefont {Burgy},
  \citenamefont {Mayr}, \citenamefont {Martin-Mayor}, \citenamefont {Moreo},\
  and\ \citenamefont {Dagotto}}]{PhysRevLett.87.277202}%
  \BibitemOpen
  \bibfield  {author} {\bibinfo {author} {\bibfnamefont {J.}~\bibnamefont
  {Burgy}}, \bibinfo {author} {\bibfnamefont {M.}~\bibnamefont {Mayr}},
  \bibinfo {author} {\bibfnamefont {V.}~\bibnamefont {Martin-Mayor}}, \bibinfo
  {author} {\bibfnamefont {A.}~\bibnamefont {Moreo}},\ and\ \bibinfo {author}
  {\bibfnamefont {E.}~\bibnamefont {Dagotto}},\ }\href
  {https://doi.org/10.1103/PhysRevLett.87.277202} {\bibfield  {journal}
  {\bibinfo  {journal} {Phys. Rev. Lett.}\ }\textbf {\bibinfo {volume} {87}},\
  \bibinfo {pages} {277202} (\bibinfo {year} {2001})}\BibitemShut {NoStop}%
\bibitem [{\citenamefont {Binder}\ and\ \citenamefont
  {Young}(1986)}]{RevModPhys.58.801}%
  \BibitemOpen
  \bibfield  {author} {\bibinfo {author} {\bibfnamefont {K.}~\bibnamefont
  {Binder}}\ and\ \bibinfo {author} {\bibfnamefont {A.~P.}\ \bibnamefont
  {Young}},\ }\href {https://doi.org/10.1103/RevModPhys.58.801} {\bibfield
  {journal} {\bibinfo  {journal} {Rev. Mod. Phys.}\ }\textbf {\bibinfo {volume}
  {58}},\ \bibinfo {pages} {801} (\bibinfo {year} {1986})}\BibitemShut
  {NoStop}%
\bibitem [{\citenamefont {Mydosh}(1993)}]{mydosh1993spin}%
  \BibitemOpen
  \bibfield  {author} {\bibinfo {author} {\bibfnamefont {J.~A.}\ \bibnamefont
  {Mydosh}},\ }\href@noop {} {\emph {\bibinfo {title} {Spin glasses: an
  experimental introduction}}}\ (\bibinfo  {publisher} {CRC Press},\ \bibinfo
  {year} {1993})\BibitemShut {NoStop}%
\bibitem [{\citenamefont {Skomski}(2008)}]{Skomski2008Magnetism}%
  \BibitemOpen
  \bibfield  {author} {\bibinfo {author} {\bibfnamefont {R.}~\bibnamefont
  {Skomski}},\ }\href
  {https://doi.org/10.1093/acprof:oso/9780198570752.001.0001} {\emph {\bibinfo
  {title} {{Simple Models of Magnetism}}}}\ (\bibinfo  {publisher} {Oxford
  University Press},\ \bibinfo {year} {2008})\BibitemShut {NoStop}%
\bibitem [{\citenamefont {Meiklejohn}\ and\ \citenamefont
  {Bean}(1957)}]{PhysRev.105.904}%
  \BibitemOpen
  \bibfield  {author} {\bibinfo {author} {\bibfnamefont {W.~H.}\ \bibnamefont
  {Meiklejohn}}\ and\ \bibinfo {author} {\bibfnamefont {C.~P.}\ \bibnamefont
  {Bean}},\ }\href {https://doi.org/10.1103/PhysRev.105.904} {\bibfield
  {journal} {\bibinfo  {journal} {Phys. Rev.}\ }\textbf {\bibinfo {volume}
  {105}},\ \bibinfo {pages} {904} (\bibinfo {year} {1957})}\BibitemShut
  {NoStop}%
\bibitem [{\citenamefont {Zener}(1954)}]{PhysRev.96.1335}%
  \BibitemOpen
  \bibfield  {author} {\bibinfo {author} {\bibfnamefont {C.}~\bibnamefont
  {Zener}},\ }\href {https://doi.org/10.1103/PhysRev.96.1335} {\bibfield
  {journal} {\bibinfo  {journal} {Phys. Rev.}\ }\textbf {\bibinfo {volume}
  {96}},\ \bibinfo {pages} {1335} (\bibinfo {year} {1954})}\BibitemShut
  {NoStop}%
\bibitem [{\citenamefont {Daalderop}\ \emph {et~al.}(1996)\citenamefont
  {Daalderop}, \citenamefont {Kelly},\ and\ \citenamefont
  {Schuurmans}}]{PhysRevB.53.14415}%
  \BibitemOpen
  \bibfield  {author} {\bibinfo {author} {\bibfnamefont {G.~H.~O.}\
  \bibnamefont {Daalderop}}, \bibinfo {author} {\bibfnamefont {P.~J.}\
  \bibnamefont {Kelly}},\ and\ \bibinfo {author} {\bibfnamefont {M.~F.~H.}\
  \bibnamefont {Schuurmans}},\ }\href
  {https://doi.org/10.1103/PhysRevB.53.14415} {\bibfield  {journal} {\bibinfo
  {journal} {Phys. Rev. B}\ }\textbf {\bibinfo {volume} {53}},\ \bibinfo
  {pages} {14415} (\bibinfo {year} {1996})}\BibitemShut {NoStop}%
\end{thebibliography}%


\begin{thebibliography}{9}%
\makeatletter
\providecommand \@ifxundefined [1]{%
 \@ifx{#1\undefined}
}%
\providecommand \@ifnum [1]{%
 \ifnum #1\expandafter \@firstoftwo
 \else \expandafter \@secondoftwo
 \fi
}%
\providecommand \@ifx [1]{%
 \ifx #1\expandafter \@firstoftwo
 \else \expandafter \@secondoftwo
 \fi
}%
\providecommand \natexlab [1]{#1}%
\providecommand \enquote  [1]{``#1''}%
\providecommand \bibnamefont  [1]{#1}%
\providecommand \bibfnamefont [1]{#1}%
\providecommand \citenamefont [1]{#1}%
\providecommand \href@noop [0]{\@secondoftwo}%
\providecommand \href [0]{\begingroup \@sanitize@url \@href}%
\providecommand \@href[1]{\@@startlink{#1}\@@href}%
\providecommand \@@href[1]{\endgroup#1\@@endlink}%
\providecommand \@sanitize@url [0]{\catcode `\\12\catcode `\$12\catcode
  `\&12\catcode `\#12\catcode `\^12\catcode `\_12\catcode `\%12\relax}%
\providecommand \@@startlink[1]{}%
\providecommand \@@endlink[0]{}%
\providecommand \url  [0]{\begingroup\@sanitize@url \@url }%
\providecommand \@url [1]{\endgroup\@href {#1}{\urlprefix }}%
\providecommand \urlprefix  [0]{URL }%
\providecommand \Eprint [0]{\href }%
\providecommand \doibase [0]{https://doi.org/}%
\providecommand \selectlanguage [0]{\@gobble}%
\providecommand \bibinfo  [0]{\@secondoftwo}%
\providecommand \bibfield  [0]{\@secondoftwo}%
\providecommand \translation [1]{[#1]}%
\providecommand \BibitemOpen [0]{}%
\providecommand \bibitemStop [0]{}%
\providecommand \bibitemNoStop [0]{.\EOS\space}%
\providecommand \EOS [0]{\spacefactor3000\relax}%
\providecommand \BibitemShut  [1]{\csname bibitem#1\endcsname}%
\let\auto@bib@innerbib\@empty
\bibitem [{\citenamefont {Kumar}\ \emph {et~al.}(2021)\citenamefont {Kumar},
  \citenamefont {Tiwari}, \citenamefont {Chauhan},\ and\ \citenamefont
  {Ghosh}}]{Kumar2021}%
  \BibitemOpen
  \bibfield  {author} {\bibinfo {author} {\bibfnamefont {B.}~\bibnamefont
  {Kumar}}, \bibinfo {author} {\bibfnamefont {J.~K.}\ \bibnamefont {Tiwari}},
  \bibinfo {author} {\bibfnamefont {H.~C.}\ \bibnamefont {Chauhan}},\ and\
  \bibinfo {author} {\bibfnamefont {S.}~\bibnamefont {Ghosh}},\ }\href
  {https://doi.org/10.1038/s41598-021-00544-8} {\bibfield  {journal} {\bibinfo
  {journal} {Scientific Reports}\ }\textbf {\bibinfo {volume} {11}},\ \bibinfo
  {pages} {21184} (\bibinfo {year} {2021})}\BibitemShut {NoStop}%
\bibitem [{\citenamefont {Dabrowski}\ \emph {et~al.}(1999)\citenamefont
  {Dabrowski}, \citenamefont {Xiong}, \citenamefont {Bukowski}, \citenamefont
  {Dybzinski}, \citenamefont {Klamut}, \citenamefont {Siewenie}, \citenamefont
  {Chmaissem}, \citenamefont {Shaffer}, \citenamefont {Kimball}, \citenamefont
  {Jorgensen},\ and\ \citenamefont {Short}}]{PhysRevB.60.7006}%
  \BibitemOpen
  \bibfield  {author} {\bibinfo {author} {\bibfnamefont {B.}~\bibnamefont
  {Dabrowski}}, \bibinfo {author} {\bibfnamefont {X.}~\bibnamefont {Xiong}},
  \bibinfo {author} {\bibfnamefont {Z.}~\bibnamefont {Bukowski}}, \bibinfo
  {author} {\bibfnamefont {R.}~\bibnamefont {Dybzinski}}, \bibinfo {author}
  {\bibfnamefont {P.~W.}\ \bibnamefont {Klamut}}, \bibinfo {author}
  {\bibfnamefont {J.~E.}\ \bibnamefont {Siewenie}}, \bibinfo {author}
  {\bibfnamefont {O.}~\bibnamefont {Chmaissem}}, \bibinfo {author}
  {\bibfnamefont {J.}~\bibnamefont {Shaffer}}, \bibinfo {author} {\bibfnamefont
  {C.~W.}\ \bibnamefont {Kimball}}, \bibinfo {author} {\bibfnamefont {J.~D.}\
  \bibnamefont {Jorgensen}},\ and\ \bibinfo {author} {\bibfnamefont
  {S.}~\bibnamefont {Short}},\ }\href
  {https://doi.org/10.1103/PhysRevB.60.7006} {\bibfield  {journal} {\bibinfo
  {journal} {Phys. Rev. B}\ }\textbf {\bibinfo {volume} {60}},\ \bibinfo
  {pages} {7006} (\bibinfo {year} {1999})}\BibitemShut {NoStop}%
\bibitem [{\citenamefont {Urushibara}\ \emph {et~al.}(1995)\citenamefont
  {Urushibara}, \citenamefont {Moritomo}, \citenamefont {Arima}, \citenamefont
  {Asamitsu}, \citenamefont {Kido},\ and\ \citenamefont
  {Tokura}}]{PhysRevB.51.14103}%
  \BibitemOpen
  \bibfield  {author} {\bibinfo {author} {\bibfnamefont {A.}~\bibnamefont
  {Urushibara}}, \bibinfo {author} {\bibfnamefont {Y.}~\bibnamefont
  {Moritomo}}, \bibinfo {author} {\bibfnamefont {T.}~\bibnamefont {Arima}},
  \bibinfo {author} {\bibfnamefont {A.}~\bibnamefont {Asamitsu}}, \bibinfo
  {author} {\bibfnamefont {G.}~\bibnamefont {Kido}},\ and\ \bibinfo {author}
  {\bibfnamefont {Y.}~\bibnamefont {Tokura}},\ }\href
  {https://doi.org/10.1103/PhysRevB.51.14103} {\bibfield  {journal} {\bibinfo
  {journal} {Phys. Rev. B}\ }\textbf {\bibinfo {volume} {51}},\ \bibinfo
  {pages} {14103} (\bibinfo {year} {1995})}\BibitemShut {NoStop}%
\bibitem [{\citenamefont {Paraskevopoulos}\ \emph {et~al.}(2000)\citenamefont
  {Paraskevopoulos}, \citenamefont {Mayr}, \citenamefont {Hartinger},
  \citenamefont {Pimenov}, \citenamefont {Hemberger}, \citenamefont
  {Lunkenheimer}, \citenamefont {Loidl}, \citenamefont {Mukhin}, \citenamefont
  {Ivanov},\ and\ \citenamefont {Balbashov}}]{PARASKEVOPOULOS2000118}%
  \BibitemOpen
  \bibfield  {author} {\bibinfo {author} {\bibfnamefont {M.}~\bibnamefont
  {Paraskevopoulos}}, \bibinfo {author} {\bibfnamefont {F.}~\bibnamefont
  {Mayr}}, \bibinfo {author} {\bibfnamefont {C.}~\bibnamefont {Hartinger}},
  \bibinfo {author} {\bibfnamefont {A.}~\bibnamefont {Pimenov}}, \bibinfo
  {author} {\bibfnamefont {J.}~\bibnamefont {Hemberger}}, \bibinfo {author}
  {\bibfnamefont {P.}~\bibnamefont {Lunkenheimer}}, \bibinfo {author}
  {\bibfnamefont {A.}~\bibnamefont {Loidl}}, \bibinfo {author} {\bibfnamefont
  {A.}~\bibnamefont {Mukhin}}, \bibinfo {author} {\bibfnamefont
  {V.}~\bibnamefont {Ivanov}},\ and\ \bibinfo {author} {\bibfnamefont
  {A.}~\bibnamefont {Balbashov}},\ }\href
  {https://doi.org/https://doi.org/10.1016/S0304-8853(99)00722-2} {\bibfield
  {journal} {\bibinfo  {journal} {Journal of Magnetism and Magnetic Materials}\
  }\textbf {\bibinfo {volume} {211}},\ \bibinfo {pages} {118} (\bibinfo {year}
  {2000})}\BibitemShut {NoStop}%
\bibitem [{\citenamefont {Hemberger}\ \emph {et~al.}(2002)\citenamefont
  {Hemberger}, \citenamefont {Krimmel}, \citenamefont {Kurz}, \citenamefont
  {Krug~von Nidda}, \citenamefont {Ivanov}, \citenamefont {Mukhin},
  \citenamefont {Balbashov},\ and\ \citenamefont {Loidl}}]{PhysRevB.66.094410}%
  \BibitemOpen
  \bibfield  {author} {\bibinfo {author} {\bibfnamefont {J.}~\bibnamefont
  {Hemberger}}, \bibinfo {author} {\bibfnamefont {A.}~\bibnamefont {Krimmel}},
  \bibinfo {author} {\bibfnamefont {T.}~\bibnamefont {Kurz}}, \bibinfo {author}
  {\bibfnamefont {H.-A.}\ \bibnamefont {Krug~von Nidda}}, \bibinfo {author}
  {\bibfnamefont {V.~Y.}\ \bibnamefont {Ivanov}}, \bibinfo {author}
  {\bibfnamefont {A.~A.}\ \bibnamefont {Mukhin}}, \bibinfo {author}
  {\bibfnamefont {A.~M.}\ \bibnamefont {Balbashov}},\ and\ \bibinfo {author}
  {\bibfnamefont {A.}~\bibnamefont {Loidl}},\ }\href
  {https://doi.org/10.1103/PhysRevB.66.094410} {\bibfield  {journal} {\bibinfo
  {journal} {Phys. Rev. B}\ }\textbf {\bibinfo {volume} {66}},\ \bibinfo
  {pages} {094410} (\bibinfo {year} {2002})}\BibitemShut {NoStop}%
\bibitem [{\citenamefont {Markovich}\ \emph {et~al.}(2014)\citenamefont
  {Markovich}, \citenamefont {Wisniewski},\ and\ \citenamefont
  {Szymczak}}]{MARKOVICH20141}%
  \BibitemOpen
  \bibfield  {author} {\bibinfo {author} {\bibfnamefont {V.}~\bibnamefont
  {Markovich}}, \bibinfo {author} {\bibfnamefont {A.}~\bibnamefont
  {Wisniewski}},\ and\ \bibinfo {author} {\bibfnamefont {H.}~\bibnamefont
  {Szymczak}}\ }(\bibinfo  {publisher} {Elsevier},\ \bibinfo {year} {2014})\
  pp.\ \bibinfo {pages} {1--201}\BibitemShut {NoStop}%
\bibitem [{\citenamefont {Meiklejohn}\ and\ \citenamefont
  {Bean}(1957)}]{PhysRev.105.904}%
  \BibitemOpen
  \bibfield  {author} {\bibinfo {author} {\bibfnamefont {W.~H.}\ \bibnamefont
  {Meiklejohn}}\ and\ \bibinfo {author} {\bibfnamefont {C.~P.}\ \bibnamefont
  {Bean}},\ }\href {https://doi.org/10.1103/PhysRev.105.904} {\bibfield
  {journal} {\bibinfo  {journal} {Phys. Rev.}\ }\textbf {\bibinfo {volume}
  {105}},\ \bibinfo {pages} {904} (\bibinfo {year} {1957})}\BibitemShut
  {NoStop}%
\bibitem [{\citenamefont {Zener}(1954)}]{PhysRev.96.1335}%
  \BibitemOpen
  \bibfield  {author} {\bibinfo {author} {\bibfnamefont {C.}~\bibnamefont
  {Zener}},\ }\href {https://doi.org/10.1103/PhysRev.96.1335} {\bibfield
  {journal} {\bibinfo  {journal} {Phys. Rev.}\ }\textbf {\bibinfo {volume}
  {96}},\ \bibinfo {pages} {1335} (\bibinfo {year} {1954})}\BibitemShut
  {NoStop}%
\bibitem [{\citenamefont {Daalderop}\ \emph {et~al.}(1996)\citenamefont
  {Daalderop}, \citenamefont {Kelly},\ and\ \citenamefont
  {Schuurmans}}]{PhysRevB.53.14415}%
  \BibitemOpen
  \bibfield  {author} {\bibinfo {author} {\bibfnamefont {G.~H.~O.}\
  \bibnamefont {Daalderop}}, \bibinfo {author} {\bibfnamefont {P.~J.}\
  \bibnamefont {Kelly}},\ and\ \bibinfo {author} {\bibfnamefont {M.~F.~H.}\
  \bibnamefont {Schuurmans}},\ }\href
  {https://doi.org/10.1103/PhysRevB.53.14415} {\bibfield  {journal} {\bibinfo
  {journal} {Phys. Rev. B}\ }\textbf {\bibinfo {volume} {53}},\ \bibinfo
  {pages} {14415} (\bibinfo {year} {1996})}\BibitemShut {NoStop}%
\end{thebibliography}%

\end{document}


\large	
\widetext
\title{Supplemental Material for:\\Origin of magnetic anisotropy in infinite layers La$_{1-x}$Sr$_{x}$MnO$_{3}$}
	
	\affiliation
	{School of Physical Sciences, Jawaharlal Nehru University, New Delhi 110067, India}
	\author{Birendra Kumar}
	\author{Harish Chandr Chauhan}
	\author{Ajay Baro}
	\author{Jyoti Saini}
	\author{Ankita Tiwari}
	\affiliation
	{School of Physical Sciences, Jawaharlal Nehru University, New Delhi 110067, India}
	\author{Mukesh Verma}
	\affiliation
	{Department of Physics, Central University of Rajasthan, Bandar Sindri 305817, India} 
	\author{Yugandhar Bitla}
	\affiliation
	{Department of Physics, Central University of Rajasthan, Bandar Sindri 305817, India}
	\author{Subhasis Ghosh}
	\email{subhasis.ghosh.jnu@gmail.com}
	\affiliation{School of Physical Sciences, Jawaharlal Nehru University, New Delhi 110067, India%
	}
\maketitle
\section{Samples preparation}
\hspace{6 mm} All the infinite layer manganites La$_{1-x}$Sr$_{x}$MnO$_{3}$ (x = 0.125, 0.175, 0.200, 0.225, 0.300 and 0.400) have been prepared by standard solid state reaction method. Highly pure La$_{2}$O$_{3}$ (Sigma Aldrich 99.99\%), SrCO$_{3}$ (Alfa Aesar 99.995\%) and MnO$_{2}$ (Alfa Aesar 99.997\%) are taken as precursors. The stoichiometric ratio of La$_{2}$O$_{3}$ (preheated at 1000$ ^{0} $C for 12h), SrCO$_{3}$ (dried in air at 150 $^{0}$C for 12h) and MnO$_{2}$ for different samples were thoroughly mixed together and ground properly to get a homogeneous mixture and then calcined in air at different temperatures ranging from 1000 $^{0}$C to 1350 $^{0}`$C followed by intermediate grinding. The calcined powder is reground and pressed into the pellet by using a hydraulic press and the pellet is then sintered at a higher temperature for different compositions as tabulated in table \ref{table1}. The final sintering process is repeated by an intermediate regrinding to get the single phase of the manganite samples \cite{Kumar2021}. The temperature and time profiles such as annealing temperature and time, and sintering temperature and time for their preparation are presented in table \ref{table1}.   
\begin{table}[htb]
	\centering
	\caption {Temperature and time profiles for samples preparation.}	
	\begin{tabular}{| c | c | c | c |c |} \hline 
		& ann. Temp  & sint. Temp & ann. time & sint. time \\
		x & ($^{0}$C)  & ($^{0}$C)& (hrs) & (hrs) \\ \hline
		0.125 & 1050  & 1350 & 24 & 72 \\ \hline
		0.175 & 1050  & 1400 & 24 & 72 \\ \hline
		0.20 & 1050  & 1400 & 24 & 72 \\ \hline
		0.225 & 1050  & 1400 & 24 & 96 \\ \hline
		0.300 & 1050  & 1400 & 24 & 96 \\ \hline
		0.400 & 1050  & 1400 & 24 & 48 \\ \hline
	\end{tabular}
	\label{table1}
\end{table}

\section{Crystalographic study}
From table \ref{Table 2}, it is clear that the crystal structure (symmetry) and crystal parameters of manganites change with doping. The structure (space group) changes from orthorhombic to rhombohedral to trigonal as follows: the structure changes from orthorombic (Pnma) to rhombohedral ({R\={3}c}) for concentration change $ 0.125 \leq x \leq 0.175 $, rhombohedral to trigonal ({R\={3}c}) for concentration change $ 0.175 \leq x \leq 0.200 $ and the trigonal ({R\={3}c}) structure remains unchanged for further increase in doping concentration $ 0.200 \leq x \leq 0.400 $. The crystal parameters also changes as follows: "a" decreases (5.5427 \AA $-$5.4872 \AA) with increase in x for the entire range $ 0.125 \leq x \leq 0.400 $, the "b (\AA), c (\AA) and V (\AA$^{3}$)" increases (parameters c and V changes very fast due to change in the crystal structure) with x for $ 0.125 \leq x \leq 0.175 $ and decreases (slow) with x for $ 0.175 \leq x \leq 0.400 $. The bond length of Mn$ - $O bond in the octahedron also changes with doping as seen from the table \ref{Table 2}. The d$ ^{\parallel} $$ _{MnO_{2}} $ denotes the Mn$ - $O bond length in MnO$ _{2} $ plane while d$ ^{\perp} $$ _{MnO_{2}} $ denotes Mn$ - $O bond length perpendicular to the MnO$ _{2} $ plane. Two different values of d$ ^{\parallel} $$ _{MnO_{2}} $ (2.265 \AA, and  1.649 \AA) and one value of d$ ^{\perp} $$ _{MnO_{2}} $ (2.001 \AA) indicates that IL-LSMO-0.125 is JT distorted while for all remaining compositions d$ ^{\parallel} $$ _{MnO_{2}} $ = d$ ^{\perp} $$ _{MnO_{2}} $ which indicates that there is no JT distortion. 
\begin{table*}
		\centering
		\caption {Room temperature structural parameters for {IL-LSMO-0.125}, {IL-LSMO-0.175}, {IL-LSMO-0.200}, {IL-LSMO-0.225}, {IL-LSMO-0.300} and {IL-LSMO-0.400} obtained from Retveild refinement.}
\begin{tabular}{| c | c | c | c |c | c | c |}
	\hline 
	Parameters & IL-LSMO & IL-LSMO & IL-LSMO & IL-LSMO & IL-LSMO & IL-LSMO \\	\hline
	x & 0.125 & 0.175& 0.200 & 0.225 & 0.300 & 0.400\\
	\hline
	Symmetry  &Ortho &Rhombo &Trigonal &Trigonal &Trigonal &Trigonal \\ \hline
	Space group &Pnma & {R\={3}c} & {R\={3}c} & {R\={3}c} & {R\={3}c} & {R\={3}c} \\	\hline
	a (\AA)  & 5.5427 & 5.53820  & 5.52610  & 5.5251  & 5.5019 & 5.4872 \\ \hline 
	b (\AA)  & 5.5093 & 5.53820 & 5.52610   & 5.5251  & 5.5019 & 5.4872 \\ \hline
	c (\AA)  & 7.7997 & 13.3740 & 13.3713  & 13.3735  & 13.3563 & 13.3611 \\ \hline
	d$ ^{\parallel} $$ _{MnO_{2}} $(\AA)  & 2.265 and 1.649  & 1.979 & 1.95986  & 1.96433  & 1.94946 & 1.94532 \\ \hline
	d$ ^{\perp} $$ _{MnO_{2}} $(\AA)  & 2.001  & 1.979 & 1.95986  & 1.96433  & 1.94946 & 1.94532 \\	\hline
	V(\AA$^3$)    & 238.1749 & 255.2432  & 353.6227 & 353.5593 & 350.1387 & 348.3983 \\	\hline
	R$_p$ (\%)    & 12.5    & 12.7       & 22.6     & 19.7    & 20.2 & 20.8 \\	\hline
	R$_{wp}$ (\%) & 18.4    & 17.1      & 26.1     & 23.1    & 23.2 & 25.2 \\	\hline
	R$_F$ (\%)    & 3.76   & 2.800      & 6.04    & 6.96   & 5.71 & 5.57 \\	\hline
	$\Xi^2$ (\%)  & 1.76    & 1.32      & 1.954     & 1.815    & 2.132 & 2.208 \\
	\hline
\end{tabular}
\label{Table 2}
\end{table*} 

\section{Transition Temperature}
\begin{figure}[h]
	\centering{}
	\includegraphics[width=12cm]{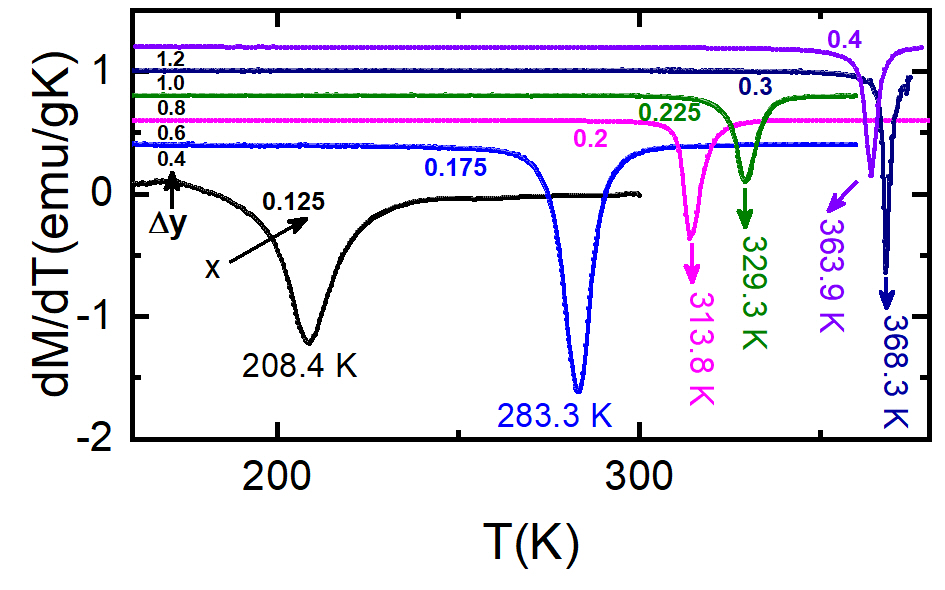}
	\caption{Doping dependent transition temperature: the transition temperature increases for $ 0.125 \leq x \leq 0.300 $ and then gets decrease for $ x = 0.400 $.}
	\label{fig1}
\end{figure}

\hspace{6 mm}
The transition temperature (T$ _{c} $) for temperature$ - $dependent (M$ - $T) magnetizations in main text, is determined as the minimum of the first derivation of M$ - $T. The corresponding $T_C$'s of IL-LSMO-$x$ have been shown in Fig. \ref{fig1}. The $T_C$ has been found to increase with Sr doping till $x=0.300$, with a decrease in $T_C$ at $x=0.400$. The trend of change in transition temperature is consistent with earlier reports \cite{PhysRevB.60.7006, PhysRevB.51.14103, PARASKEVOPOULOS2000118}. 

\section{Resistivity}
\hspace{6 mm} Based on the temperature-dependent resistivity measurements of IL-LSMO-$x$ (x = 0.125-0.400) presented in the referenced studies A. Urushibara et al.\cite{PhysRevB.51.14103} and B. Dabrowski et al.\cite{PhysRevB.60.7006}, the electrical and magnetic phases of these manganites can be understood as follows: (1) IL-LSMO-$x$ materials exhibit a metal-insulator transition (MIT) for certain compositions, namely x = 0.125, 0.130, 0.135, 0.140, 0.145, 0.150, 0.155, 0.160, 0.165, 0.170, 0.175, 0.200, and 0.300. The transition temperatures (T$ _{MIT} $) for these compositions are observed to be around 203 K, 210 K, 219 K, 225 K, 332 K, 241 K, 245 K, 254 K, 274 K, 280 K, 295 K, 313 K, and 370 K, respectively. These T$ _{MIT} $s are in consistent with the T$ _{c} $s observed from magnetization measurements. (2) IL-LSMO-$x$ materials exhibit insulating behavior below specific temperatures (T$ ^{*} $) for each composition. The T$ _{MIT} $s and corresponding temperature ranges are as follows: 168 K (168-203 K), 175 K (175-210 K), 180 K (180-219 K), 191 K (191-225 K), 200 K (200-232 K), 190 K (190-241 K), 176 K (176-246 K), 143 K (143-254 K), 127 K (127-261 K), 95 K (95-274 K), and 42 K (42-284 K) for x = 0.125, 0.130, 0.135, 0.140, 0.145, 0.150, 0.155, 0.160, 0.165, 0.170, and 0.175, respectively. (3) IL-LSMO-$x$ materials exhibit insulating behavior above their respective T$ _{MIT} $s, except for $x = 0.300$ and $0.400$, which shows metallic behavior across the entire temperature range.

The observed electrical behavior of IL-LSMO-$x$ manganites can be explained using a model based on magnetic clusters or small domains, specifically ferromagnetic (FM) clusters induced by double-exchange (DE) in the background of antiferromagnetic (AFM) regions induced by superexchange (SE). The explanation is as follows:
(1) for $ 0.125 \leq x < 0.200 $, below the T$ ^{*} $s as mentioned earlier, there are non-overlapping FM clusters in the background of AFM, leading to an insulating behavior. In the teperature range T$ ^{*} $ $<$ T $ \leq $T$ _{MIT} $, the FM clusters start to overlap, resulting in a conducting nature. Above the corresponding T$ _{MIT} $s, the clusters or spin correlations break, leading to an insulating behavior.
(2) For compositions $x = 0.200, 0.225, 0.300, and 0.400$, the FM clusters or domains overlap throughout the entire temperature range below their respective T$ _{MIT} $'s or T$ _{c} $'s, resulting in a conducting behavior. Therefore, x = 0.200 can be considered the critical composition where IL-LSMO manganite exhibits metallic behavior at all temperatures below the T$ _{MIT} $.
Thus, the electrical nature of IL-LSMO manganites can be explained using the cluster model, where the presence and overlapping of FM and AFM clusters determine the conducting or insulating behavior of the material.

\begin{figure}[h]
	\centering{}
	\includegraphics[width=\linewidth]{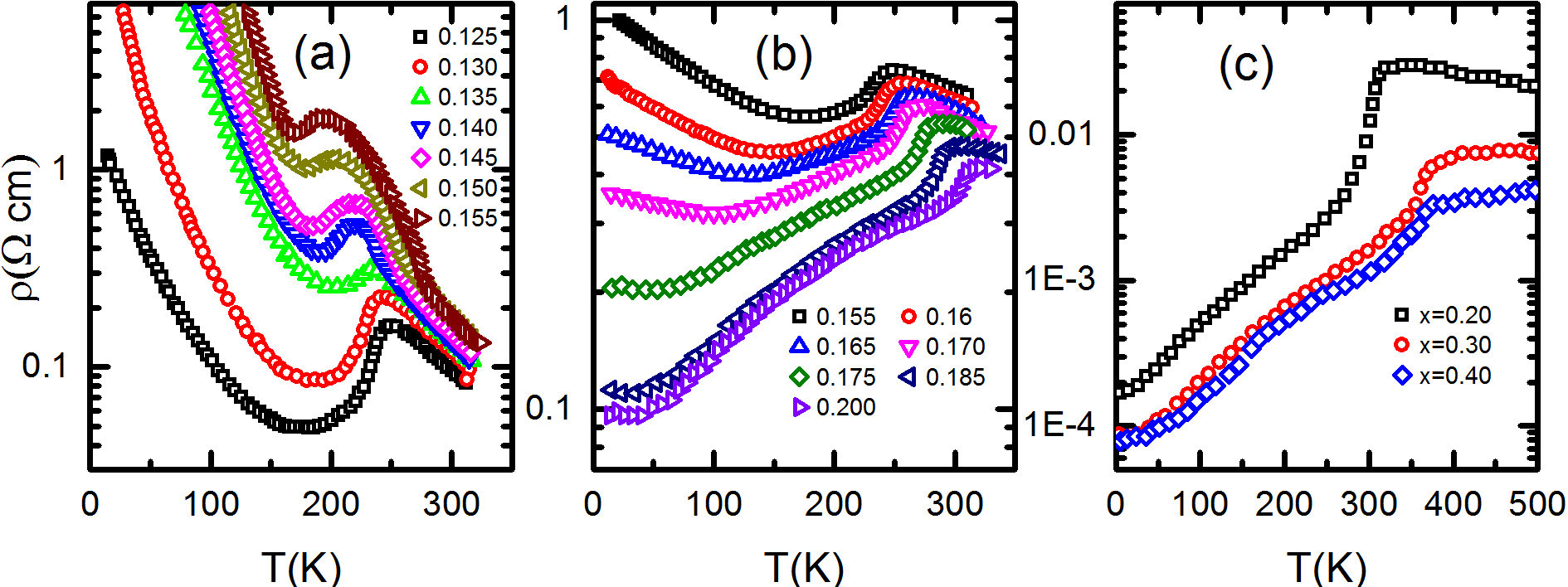}
	\caption{Temperature dependent resistivity of infinite layer manganites IL-LSMO-$ x $ for the concentration range: (a) $ x $ = 0.125-0.155, (b) $ x $ = 0.155-0.200 and (c) $ x $ = 0.200-0.400.}
	\label{fig2}
\end{figure}
 
 \section{Phase Diagram}
\begin{figure}[h]
	\centering{}
	\includegraphics[width=\linewidth]{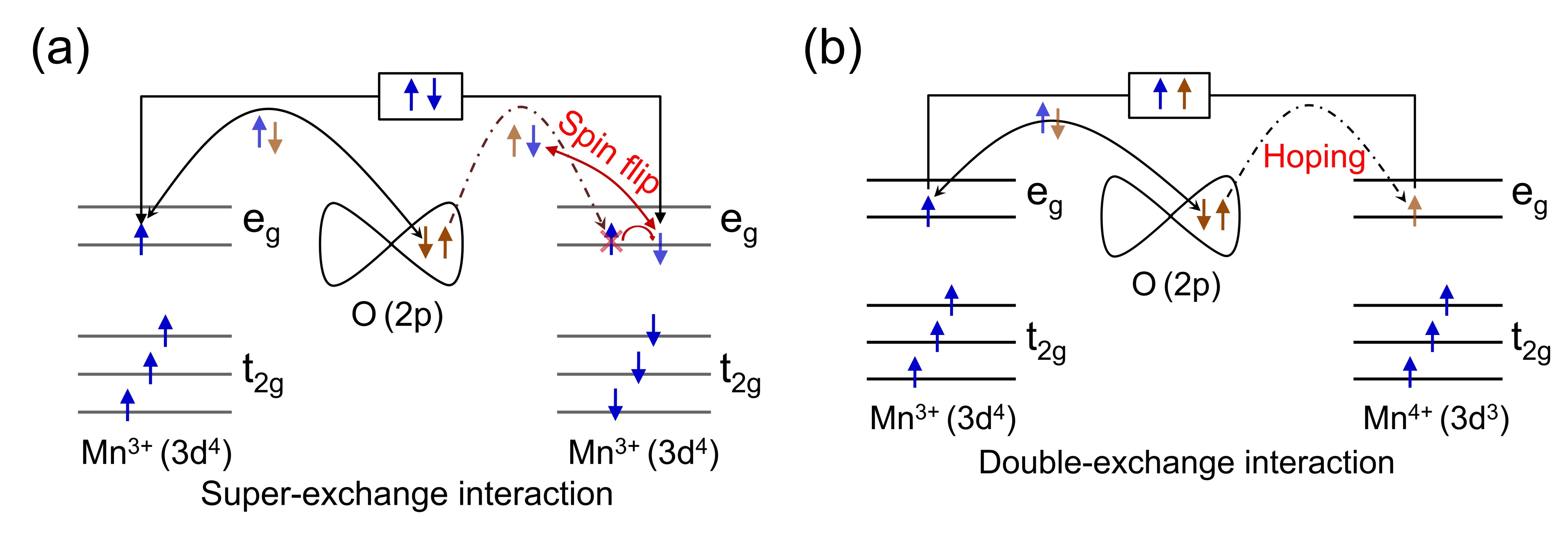}
	\caption{Schematics of (a) superexchange (Mn$^{3+}$-O$^{2-}$-Mn$^{3+}$) and (b) double$ - $exchange (Mn$^{3+}$-O$^{2-}$-Mn$^{4+}$) interactions.}
	\label{fig3}
\end{figure}

\hspace{6 mm}Further, from the transition temperatures observed from derivatives of $ dM/dT $ (Fig. 1 (b) of the main text) and electrical nature for manganites ($ 0.125 \leq x \leq 0.400 $) and with the help of J. Hemberger et al\cite{PhysRevB.66.094410} possible phase diagram can be constructed as follows. The parent compound LaMnO$_{3}$ (IL-LSMO-0.0) is an antiferromagnetic Mott-insulator. Markovich et al.\cite{MARKOVICH20141} suggested that super exchange (SE) interaction between Mn$^{3+}$ ions via O$^{2-}$ is responsible for antiferromagnetic interaction in this compound as shown in Fig. \ref{fig3} (a), where the electron virtually hops from p orbital of oxygen to e$_{g}$ state of 3d orbital in Mn$^{3+}$ (3d$^{4}$) ion (i.e. actual hoping of electron between 3d orbital of Mn$^{3+}$ ions is not allowed, leading to Mott insulator) making AFM configuration to LaMnO$_{3}$. Therefore, in LaMnO$_{3}$ the transition temperature from antiferromagnetic phase to paramagnetic phase is dependent on strength of antiferromagnetic interaction and distribution of local magnetic ions.
Since, local magnetization in antiferromagnetic materials is generally found to be in form of magnetic superstructure (two different magnetic structure of opposite orientation of spins) so their transition temperature is low. Hence, the transition temperature for IL-LSMO-0.0 is around 122 K. When LaMnO$_{3}$ is doped with Sr, the concentration of Mn$^{4+}$ ions increases with increase in sr doping x resulting in DE interaction which leads to ferromagnetic interaction. The conducting nature of doped manganites, IL-LSMO-x for $ 0.0 \leq x \leq 1.0 $, increases with Sr doping. In the mechanism of DE, an electron of e$_{g}$ state in Mn$^{3+}$ (3d$^{4}$) ion hops through O$^{2-}$ to nearest neighbor unoccupied e$_{g}$ state of Mn$^{4+}$ (3d$^{3}$) ion following Hund's rule of coupling having maximum quantum number, which leads to parallel spin orientation of spins in e$_{g}$ state and t$_{2g}$ state of Mn$^{3+}$ ions i.e. orientation of spin during hoping remains unchanged leading to parallel spin orientation in both Mn$^{3+}$ and Mn$^{4+}$ ions as shown in Fig. \ref{fig3} (b). Thus, the FM and conducting nature increases with doping composition for x $ \leq 0.4 $. As a result, the transition temperature increases with the increase in composition x (upto $\approx$ 0.4) and then its start decreasing due to over-doping of Sr for $ x \geq 0.4 $. The antiferromagnetic contribution again starts increasing due to SE interaction between pairs of any two Mn$^{3+}$ ions and pairs of any two Mn$^{4+}$ ions. Hence, the dominance of magnetic interaction in IL-LSMO-x ($0.0 < x < 1.0$) changes as follows: initially for x $\approx$ 0.0 the contribution of SE is dominating and DE is negligible. 
\begin{figure}[h]
	\centering{}
	\includegraphics[width=\linewidth]{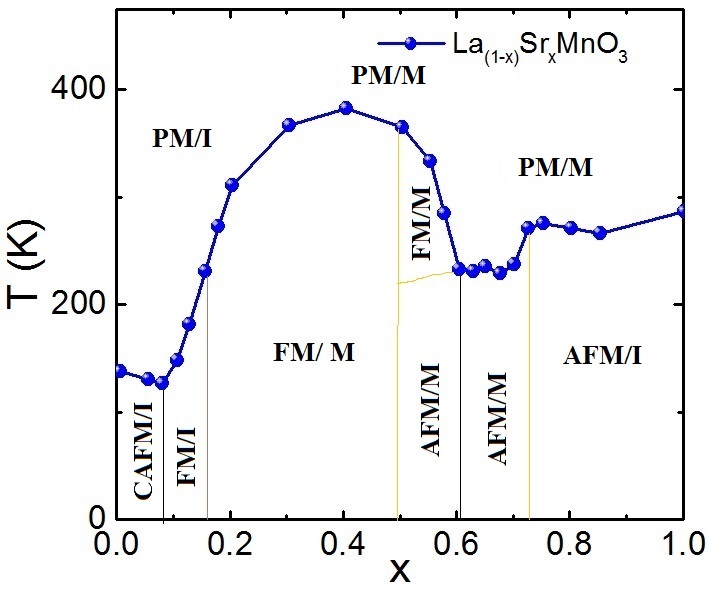}
	\caption{Phase diagram of infinite layer manganites IL-LSMO-x for $ 0.0 \leq x \leq 1.0 $, which the collective information of all possible magnetic states, electrical states, transition temperature and etc, with doping concentration. This phase diagram have been constructed based on J. Hemberger et al\cite{PhysRevB.66.094410}. I denotes Mott insulator.}
	\label{fig4}
\end{figure}
As x increases the contribution of SE decreases while DE increases leading to dominating DE over SE upto x $\approx$ 0.4. Above x = 0.4, the contribution of SE again increases. Therefore, transition temperature first increases, ($0.0 < x < 1.0$ ) and then decreases with doping. The phase diagram for IL-LSMO-x ($0.0 < x < 1.0$) is shown in Fig. \ref{fig4}\cite{PhysRevB.66.094410}. From Fig. \ref{fig4}, it is clear that for the compositions x $\leq$ 0.1 and x $\geq$ 0.6, the transition temperatures are very close to equal that may be due to dominating SE interaction over DE interaction. However, the transition temperature or say magnetic behavior of the manganites are highly sensitive to Sr doping for the composition range $0.1 \leq x \leq 0.6$, which may be due to dominating/competing DE interaction over/with SE interaction. Therefore, it is desirable to study manganites IL-LSMO-x with composition $ 0.1 < x < 0.6 $. We have selected the infinite layer Sr doped manganites La$_{1-x}$Sr$_{x}$MnO$_{3}$ for the composition range $0.125 \leq x \leq 0.4$. The transition temperatures for IL-LSMO-x, with x = 0.125, 0.175, 0.255, 0.255, 0.300 and 0.400 have been shown in Fig. \ref{fig1} which shows increase of T$ _{C} $ with composition x. This is consistence with the phase diagram in Fig. \ref{fig4}. 

\section{Magnetic anisotropy}
\hspace{6 mm} The magnetic anisotropy of magnetic materials depends on saturation magnetization (M$_{S}$) and coercive field H$_{C}$ as follows\cite{PhysRev.105.904, PhysRev.96.1335, PhysRevB.53.14415}:
\begin{equation}
E_{a} \approx K_{an}Sin^{2}\theta
\label{Kan}
\end{equation} 
When external field is applied on the sample, the Zeeman term can be written as [MIT]
\begin{equation}
E_{z} = \mu_{o}HM_{s}cos(\phi-\theta)
\label{E}
\end{equation} 
where $\mu_{o}$H is applied magnetic field, $\phi$ is the angle between external field and easy axis, $\theta$ is the angle between magnetization and easy axis. Thus, the total energy will be sum of E$_{a}$ \ref{Kan} and E$_{z}$ \ref{E}, resulting\\
\begin{equation}
E = E_{a} + E_{z} = K_{an}sin^{2}\theta + \mu_{o}HM_{s}cos(\phi-\theta)
\label{Kan1}
\end{equation} 
Magnetic field can be applied in any direction but generally, two special directions are preferred: field perpendicular to easy axis, and field parallel to the easy axis. \\
$\textbf{(1) Hard axis magnetization} $ \\
For $\phi$ = $\frac{\pi}{2}$, Eq. \ref{Kan1} becomes 
\begin{equation}
E = K_{an}sin^{2}\theta + \mu_{o}HM_{s}sin\theta
\label{E1}
\end{equation}  
Minimum value of $\theta$ for which the energy minimizes, can be determined by taking derivative of Eq. \ref{E1} w.r.t $\theta$, 
\begin{equation}
\dfrac{dE}{d\theta} = (2K_{an}sin\theta + \mu_{o}HM_{s})cos\theta=0 \implies \theta = \pm \frac{\pi}{2} + \pi n
\end{equation}
\begin{equation}
\dfrac{d^{2}E}{d\theta^{2}} = -2K_{an}sin^{2}\theta - \mu_{o}HM_{s}sin\theta + 2K_{an}cos^{2}\theta > 0
\end{equation}  
For $\theta$ = $\frac{\pi}{2}$ $\implies $ $\mu_{o}$H $<$ - $\dfrac{2K_{an}}{M_{s}}$ \\
For $\theta$ = -$\frac{\pi}{2}$ $\implies $ $\mu_{o}$H $>$ - $\dfrac{2K_{an}}{M_{s}}$ \\
The above conditions can be satisfied if:
\begin{equation}
H_{c} = \frac{2K_{an}}{M_{s}} \implies K_{an} = \frac{M_{s}H_{c}}{2}
\label{Kan2}
\end{equation}

$\textbf{(2) Easy axis magnetization} $ \\
For $\phi$ = 0, Eq. \ref{Kan1} becomes 
\begin{equation}
E = K_{an}sin^{2}\theta + \mu_{o}HM_{s}cos\theta
\end{equation}  
Minimum value of $\theta$ is obtained as follows:

\begin{equation}
\dfrac{dE}{d\theta} = (2K_{an}cos\theta - \mu_{o}HM_{s})sin\theta=0 \implies \theta = \pi n
\end{equation}
\begin{equation}
\dfrac{d^{2}E}{d\theta^{2}} = 2K_{an}cos^{2}\theta - \mu_{o}HM_{s}cos\theta - 2K_{an}sin^{2}\theta > 0
\end{equation}  
For $\theta$ = 0 $\implies $ $\mu_{o}$H $<$ - $\dfrac{2K_{an}}{M_{s}}$ \\
For $\theta$ = -$\pi$ $\implies $ $\mu_{o}$H $>$ - $\dfrac{2K_{an}}{M_{s}}$ \\
The above conditions can be satisfied if:
\begin{equation}
H_{c} = \frac{2K_{an}}{M_{s}} \implies K_{an} = \frac{M_{s}H_{c}}{2}
\label{Kan2-1}
\end{equation}
Thus, one can study the doping dependent change in magnetic anisotropy by using experimentally determined M$_{s}$ and H$_{c}$ as has been discussed in the main article (text). 

\bibliographystyle{apsrev4-2}
\bibliography{Ref_Sup}